\documentclass[epj,nopacs]{svjour}

\usepackage{graphicx}% Include figure files
\usepackage{color}
\usepackage{dcolumn}% Align table columns on decimal point
\usepackage{bm}% bold math
\usepackage{amsmath}
\usepackage{amssymb}
\usepackage{mathrsfs}
\usepackage{float}
\usepackage[caption = false]{subfig}  
\usepackage{mathrsfs}

\newcommand{\vv}{{\bf v}}

\newcommand{\vu}{{\bf u}}

\newcommand{\vx}{{\bf x}}
\newcommand{\va}{{\bf a}}

\newcommand{\bc}{\begin{center}}
\newcommand{\ec}{\end{center}}

\newcommand{\be}{\begin{equation}}
\newcommand{\ee}{\end{equation}}
\newcommand{\bea}{\begin{eqnarray}}
\newcommand{\eea}{\end{eqnarray}}

\newcommand{\euler}{e}

\usepackage{fancyhdr}
\begin{document}

\title{Anomalous dispersion in correlated porous media: A coupled
  continuous time random walk approach}

\author{Alessandro Comolli \inst{1,2,3} \and Marco Dentz \inst{1,3}}
\institute{Institute of Environmental Assessment and Water Research
  (IDAEA-CSIC), Barcelona, Spain \and Department of Civil and Environmental
  Engineering, Technical University of Catalonia
  (UPC), Barcelona, Spain \and Associated Unit: Hydrogeology Group
  (UPC-CSIC)}

\titlerunning{Anomalous dispersion in correlated porous media}

\abstract{We study the causes of anomalous dispersion in Darcy-scale
  porous media characterized by spatially heterogeneous hydraulic
  properties. Spatial variability in hydraulic conductivity leads to spatial 
  variability in the flow properties through Darcy's law and thus
  impacts on solute and particle transport. We consider
  purely advective transport in heterogeneity
  scenarios characterized by broad distributions of heterogeneity
  length scales and point values. Particle transport is characterized
  in terms of the stochastic properties of equidistantly sampled Lagrangian
  velocities, which are determined by the flow and conductivity
  statistics. The persistence length scales of flow and transport velocities are imprinted
  in the spatial disorder and reflect the distribution of
  heterogeneity length scales. Particle transitions over the
  velocity length scales are kinematically coupled with the transition
  time through velocity. We show that the average particle motion follows a
  coupled continuous time random walk (CTRW), which is fully
  parameterized by the distribution of flow velocities and
  the medium geometry in terms of the heterogeneity length scales.  
The coupled CTRW provides a systematic framework for the investigation of the
origins of anomalous dispersion in terms of heterogeneity correlation
and the distribution of heterogeneity point values.  We derive
analytical expressions for the asymptotic scaling of the moments of
the spatial particle distribution and first arrival
time distribution (FATD), and perform numerical particle tracking
simulations of the coupled CTRW to capture the full average transport
behavior. Broad distributions of heterogeneity point values and
lengths scales may lead to very similar dispersion behaviors in terms
of the spatial variance. Their mechanisms, however are very different,
which manifests in the distributions of particle positions
and arrival times, which plays a central role for the
prediction of the fate of dissolved substances in heterogeneous
natural and engineered porous materials.}

\authorrunning{Comolli and Dentz}
\maketitle

%--------------------------------------------------------------------------------------------------
\section{Introduction}
%--------------------------------------------------------------------------------------------------
%
Large scale transport in disordered media generally exhibits
non-Fickian features that cannot be captured by models based on the
advection-dispersion equation (ADE) with constant drift and dispersion
coefficients. Non-Fickian or anomalous transport characteristics
have indeed been found ubiquitously in natural and engineered
systems~\cite{BG1990,Klafter2005},
including transport of charge carriers in amorphous solids
\cite{SL73.1,ScherMontroll75}, photon transport in atomic vapors
\cite{chevrollier2010} and in L\'evy glasses \cite{barthelemy2008l},
animal foraging patterns \cite{Viswa} and
human motion \cite{Brown2006}, diffusion in living
cells \cite{Yu2009,Jeon2016,Massignan2014}, and contaminant transport in geological
formations~\cite{Berkowitz2006}.  

In this paper, we focus on solute and particle transport in Darcy-scale
heterogeneous porous media, whose applications range from solute transport
in fractured and porous geological media~\cite{Bear:1972} to chromatography and
chemical engineering~\cite{brenner:book}. Spatial heterogeneity in the physical and
chemical medium properties lead to anomalous transport behaviors
characterized by non-linear growth of variance of particle
displacements, non-Gaussian particle distributions and early and
late particle arrivals
\cite{GWR92,Cushman-1993-Nonlocal,Haggerty1995,BerkowitzScher95,l:Cvetkovic96,BS1997,Carrera1998,Haggerty2000,Willmann2008,Berkowitz2006,LBDC2008:2,Cvetkovic2014}. 
The sound understanding of these phenomena is of crucial importance for applications ranging from geological storage of nuclear waste, carbon dioxide sequestration in geological formations, geothermal energy exploration, to name a few. 
The heterogeneity impact on large scale transport through heterogeneous media has been 
quantified using stochastic-perturbative approaches to quantify macrodispersion coefficients
\cite{DAG1984,GA1983,Rubin-2003-Applied}, as well as non-local constitutive
theories \cite{CHG1994,guadagnini1999}, fractional advection-dispersion equations
\cite{Meerschaert-1999-Multidimensional,Benson2000,ZhangBenson,benson2013}, multi-rate mass transfer models 
\cite{Haggerty1995,Carrera1998,Willmann2008}, time domain random walks~\cite{l:Cvetkovic96,Delay:et:al:2005,BenkePainter2003,Fiori2007,,Cvetkovic2014,russian2016,noetinger2016} and continuous time random walks (CTRW)~\cite{BS1997,hatanohatano,DCSB2004,Berkowitz2006,LBDC2008:2} to account for anomalous transport features in spatial distributions and arrival times.  

Continuous time random walks~\cite{MW1965} provide a natural approach to dispersion in disordered media, for which 
transport properties such as particle velocities and retention are persistent in space. Thus, particle motion can be characterized through a series of spatial and temporal transitions, which are determined by the statistical medium properties. Independence of subsequent space and time increments requires that the spatial disorder is sampled efficiently by the microscopic particle motion, this means, particles should in average explore ever new aspects of the disorder~\cite{BG1990}. This is the case for purely diffusive motion in $d > 2$ dimensional disordered media~\cite{BG1990,DRG2016}, and for biased motion in random media in any dimension. Thus, the CTRW approach has been used for the modeling of anomalous dispersion for a broad range of particle motions in random media~\cite{Metzler2000,Klafter2005,Berkowitz2006,BarkaiPT2012,Kutner2017,Shlesinger2017,} starting
with the pioneering work of \textit{Scher and Lax} \cite{SL73.1} that quantifies the anomalous motion of charge carriers in amorphous solids. 

Here we focus on solute and particle transport in heterogeneous porous media. \textit{Saffman}~\cite{saffman1959} used an approach very similar to CTRW for the quantification of pore-scale particle motion and the derivation for dispersion coefficients. Anomalous transport due to pore scale flow heterogeneity has been modeled with CTRW approaches based on detailed numerical simulations~\cite{Bijeljic2006,LeBorgne2011,Bijeljic2011,bijeljic2013,deanna2013,Kang2014,Gjetvaj2015} and laboratory scale experiments~\cite{Holzner2015}. These approaches are based on the property that particle velocities are persistent over a characteristic pore scale such that the transition time is given kinematically by the transition length and the flow velocity~\cite{Dentz2016}. The work of \textit{Berkowitz and Scher}~\cite{BS1997} has used the CTRW approach for the characterization of anomalous solute dispersion in fractured media, the work by \textit{Hatano and Hatano}~\cite{hatanohatano} for the interpretation of solute breakthrough curves in laboratory scale flow and transport experiments through columns filled with porous material. The CTRW and the related time-domain random walk (TDRW) approach~\cite{Cvetkovic2014,noetinger2016} have been used to model non-Fickian and anomalous transport features in Darcy-scale heterogeneous porous media~\cite{Berkowitz2006,noetinger2016} under uniform and non-uniform flow conditions~\cite{Kang2015,Dentzetal2015}. 
Again, the impact of advective heterogeneity is quantified through kinematic coupling of the transition length and time via the flow velocity. In this context, the CTRW has been coupled with spatial Markov models for the evolution of particle velocities along streamlines~\cite{LBDC2008:1,LBDC2008:2,Kang2011} in order to capture correlation effects of subsequent velocities and to model the impact of the initial velocity distributions on solute transport~\cite{Dentz2016,Kang2017}. Also the impact of solute retention due to mass transfer between mobile and immobile zones owing to physical or chemical interactions between the transported particle and the medium has been modeled by different CTRW approaches~\cite{Dentz2003,Margolin:et:al:2003,DB2005,DCa:2009,BensonMeer2009,Dentzetal2012,Gjetvaj2015,Comolli2016}. 

We investigate here two particular aspects of transport through heterogeneous porous media, namely disorder correlation and disorder distribution, which both can give rise to anomalous dispersion in disordered media~\cite{BG1990}. Distribution versus correlation induced anomalous transport was studied for biased particle motion in $d = 1$ dimensional media characterized by spatially varying retention properties~\cite{DentzBolster2010}. Here we focus on advective particle motion through Darcy scale porous media characterized by spatially variable hydraulic conductivity. Hydraulic conductivity is the central material property for the understanding of flow and transport in porous media. It varies in natural media over up to 12 orders of magnitude~\cite{Bear:1972}.  For Darcy scale porous and fractured media, the distribution of hydraulic conductivity is mapped onto the flow velocity via the Darcy equation~\cite{Bear:1972}. For low hydraulic conductivities, which are of particular relevance for the occurrence of anomalous transport, the conductivity has been shown to be proportional to the magnitude of the Eulerian flow velocities~\cite{Fiori2007,tyukhova2016}, which in turn can be related to the particle velocities~\cite{Dentz2016}. We consider porous media characterized by strong spatial correlation of hydraulic conductivity and thus flow velocity, expressed by a distribution of characteristic persistence scale, as well as broad heterogeneity point distributions. The objective is to derive the governing equation for the average particle motion and investigate and quantify the impacts of heterogeneity distribution and heterogeneity correlation on average particle transport in terms of spatial particle distributions, arrival times and dispersion. 

This paper is organized as follows. The flow and transport model as well as the porous media model are discussed in Sect. \ref{sectionModel}. Section~\ref{coarsegraining} derives a coupled CTRW model for average particle motion based on coarse-graining of the microscopic equations of motion and ensemble averaging.  Section~\ref{section:transpbehaviour} uses the derived model
to investigate the transport behavior in three different disorder scenarios that are characterized by  distribution-induced anomalous transport, correlation-induced anomalous transport and anomalous transport induced by both distribution and correlation. For each scenario, we derive the asymptotic scalings of the moments and the first arrival time distributions and we perform numerical
simulations. 

%--------------------------------------------------------------------------------------------------
\section{Physical model} \label{sectionModel}
%--------------------------------------------------------------------------------------------------
In the following, we present the basics of flow and advective transport in Darcy scale heterogeneous porous media 
and specify the statistical properties of the heterogeneous media model under consideration. 
%%%%%%%%%%%%%%%%%%%%%%%%%%%%%%%%%%%%%%%%%%%%%%%%%%%%
\subsection{Flow and transport in porous media} \label{FlowTransp}
%%%%%%%%%%%%%%%%%%%%%%%%%%%%%%%%%%%%%%%%%%%%%%%%%%%%
Flow through heterogeneous porous media is described by the Darcy equation~\cite{Bear:1972} for the Eulerian flow field $\vu(\vx)$
\begin{align}\label{darcy}
\vu(\vx) = -K(\vx) \nabla h(\vx), 
\end{align}
where $K(\vx)$ is hydraulic conductivity and $h(\vx)$ is hydraulic
head. We assume that both medium and fluid are incompressible and thus
$\nabla \cdot \vu(\vx) = 0$, which implies
\begin{equation}
\label{flow}
\nabla K(\vx) \cdot \nabla h(\vx) + K(\vx)\nabla^2 h(\vx) = 0.
\end{equation}
The position vector here is where $\vx = (x,y,z)^\top$. The absolute Eulerian velocity is denoted 
by $v_e(\vx) = \lVert \vu(\vx) \rVert$, where $\lVert \cdot \rVert$ denotes the $\ell^2$ norm. 
The spatially varying hydraulic conductivity depends both on the  medium and fluid properties. The fluid properties are constant here, thus it expresses the permeability of the  porous medium. The hydraulic conductivity is modeled as a stationary and ergodic spatial random field~\cite{Christakos,Rubin-2003-Applied}, whose statistical properties are discussed in the next section. The stochasticity of $K(\vx)$ is mapped onto the flow velocity through Eq. ~\eqref{darcy}. Ergodicity implies that the probability density function (PDF) $p_e(v)$ of velocity point values $v_e(\vx)$ sampled in space is equal to ensemble sampling, $p_e(v) = \langle \delta[v - v_e(\vx)] \rangle$, where the angular brackets denote the disorder average and $\delta(\cdot)$ denotes the Dirac delta-distribution. 
We consider in the following a global hydraulic head gradient aligned with the $x$-direction, which drives the flow through the porous medium. 

We consider here purely advective transport, which is described by the
advection equation
\begin{align}
\label{adest}
\frac{d \vx(t)}{d t} = \vu[\vx(t)]. 
\end{align}
For steady flows, streamlines and particle trajectories are identical. 
The distance $s(t)$ a particle covers along a streamline is given by 
\begin{align}
\frac{ds(t)}{dt} = v_t(t), && v_t(t) = \lVert\vu[\vx(t)]\rVert. 
\end{align}
We perform now a change of variables from $t \to s$ according to~\cite{DentzBolster2010,Comolli2016,Dentz2016}  
\begin{subequations}
\label{ades}
\begin{align}
\label{langevinds}
d t = \frac{ds}{v_s(s)}, && v_s(s) = \lVert \vu[\vx(s)] \rVert 
\end{align}
such that the advection equation~\eqref{adest} transforms to
\begin{align}
\label{xst}
\frac{d \vx(s)}{d s} = \frac{\vv_s(s)}{v_s(s)}, && \vv_s(s) = \vu[\vx(s)]. 
\end{align}
\end{subequations}
The particle velocities $v_s(s)$ are sampled equidistantly along streamlines as opposed to the classical definition of 
Lagrangian velocities given by $v_t$ which are sampled isochronally along streamlines~\cite{Dentz2016}. 
We refer to the point probability density function (PDF) $p_s(v)$ of velocities $v_s(s)$ along streamlines as the s-Lagrangian velocity PDF. It is related to the Eulerian velocity PDF $p_e(v)$ through flux-weighting as \cite{Dentz2016}
\begin{equation}
 p_s(v) = \frac{v}{\langle v_e \rangle} p_e(v), 
\label{Eul_sLag}
\end{equation}
where $\langle v_e \rangle$ is the mean Eulerian velocity, see also Appendix \ref{EusLag}.
As initial condition, we consider here a flux-weighted particle injection, this means the number of particles 
is proportional to flow velocity at the injection point. Thus, the initial distribution of particle velocities is equal to~\eqref{Eul_sLag}.  
%%%%%%%%%%%%%%%%%%%%%%%%%%%%%%%%%%%%%%%%%
\subsection{Disorder model} \label{disorder}
\begin{figure}[h]
\begin{center}
\includegraphics[width=.45\textwidth]{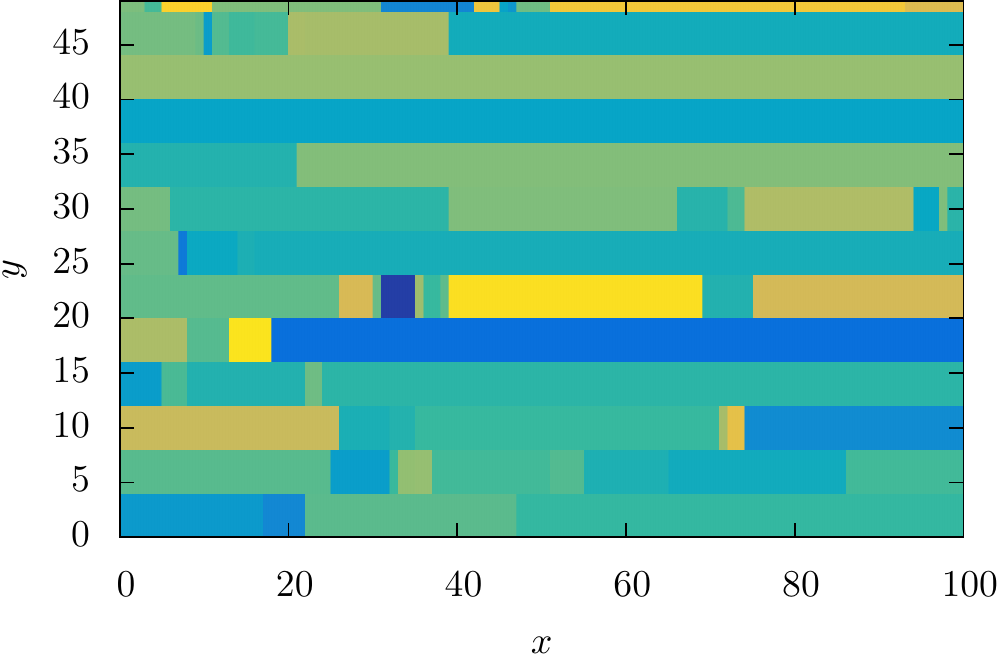} 
\includegraphics[width=.45\textwidth]{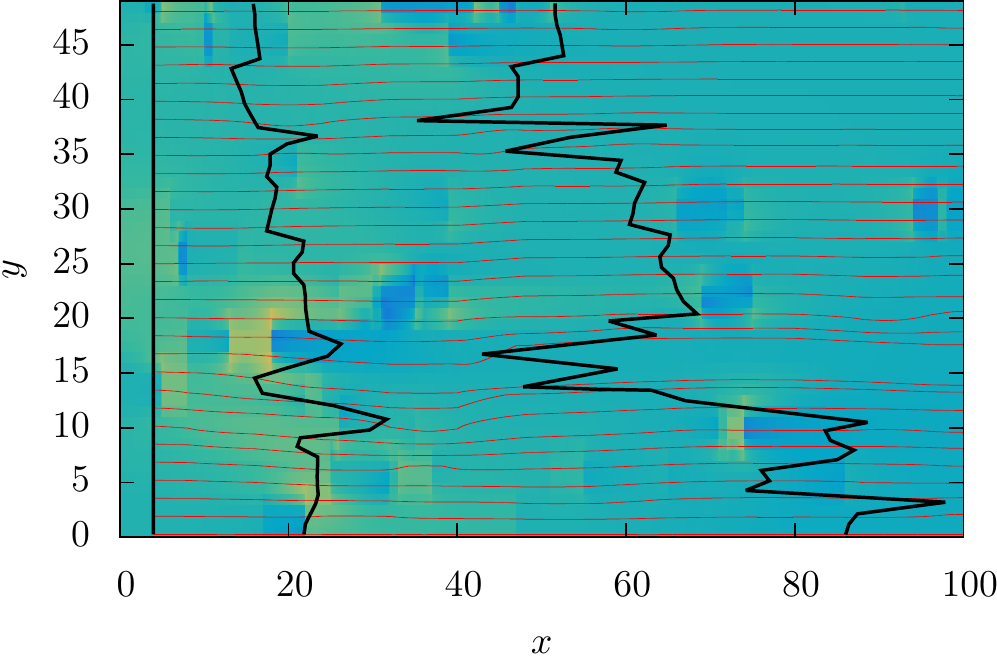}
\caption{(Top panel) Spatial distribution of hydraulic conductivity
  $K(\vx)$.  (Bottom panel) Spatial distribution of the corresponding
  Eulerian velocity $v_e(\vx)$. Dark blue denotes the lowest, yellow
  the highest values of conductivity and velocity magnitude,
  respectively. Red lines represent the streamlines and the black line shows the dispersion of particles at different times, which are injected along a line. The velocity field is obtained by solving \ref{darcy} using finite volumes with a prescribed head gradient at the vertical boundaries and no-flow conditions at the bottom and top boundaries.}
\label{illustra}
\end{center}
\end{figure}

%%%%%%%%%%%%%%%%%%%%%%%%%%%%%%%%%%%%%%%%%
We consider random media in which the hydraulic conductivity is
spatially distributed in a geometry of bins or voxels of constant
height $h_0$ and width $d_0$ and variable length $\ell$. 
We assume that the properties of the medium are constant within a
bin. Thus, we assign to the $i$-th bin the conductivity $K_i$, which
is distributed according to $p_K(K)$. The bin length $\ell$ is
distributed according to $p_\ell(\ell)$. Figure~\ref{illustra}
illustrates the heterogeneity organization and the distribution of the
velocity magnitude $v_e(\vx)$. We observe that the spatial
organization of $v_e(\vx)$ is similar to the distribution of
$K(\vx)$. In fact the relation between velocity magnitude and
hydraulic conductivity is obtained from~\eqref{darcy} as 
\begin{align}
\label{darcye}
v_e(\vx) = K(\vx) \lVert \nabla h(\vx) \rVert.
\end{align}
This means that, for an approximately constant hydraulic head gradient,
velocity magnitude and hydraulic conductivity are directly
proportional. In fact, for stratified media, this means media
characterized by infinitely long bins, the head gradient is constant
and the streamlines are parallel. Here, the streamlines are not
parallel because of fluid mass conservation as expressed by $\nabla
\cdot \vu(\vx) = 0$. Nevertheless, within a bin of constant
conductivity, the streamlines are approximately parallel as
illustrated in Fig.~\ref{illustra}. Locally, within a bin,
conductivity is constant and thus, the flow equation~\eqref{flow}
implies that the head gradient is constant. In fact, the flow field inside
a bin can be approximated by the solution for an isolated
inclusion~\cite{Eames1999,Fiori2007,Cvetkovic2014}. This implies specifically
that for small conductivities $v_e(\vx) \propto K(\vx)$, which is what
we see in Fig.~\ref{illustra}. It has been observed in numerical simulations of Darcy
scale flow that the PDF of the velocity magnitude and the PDF of
hydraulic conductivity are proportional at small values    
\cite{Edery2014,tyukhova2016,Hakoun2017}.  
Note that this local relation does not
violate fluid mass conservation because it concerns the velocity
magnitude $v_e(\vx)$ and not $\vu(\vx)$. Furthermore, the flux-weighting relation~\eqref{Eul_sLag} between the
PDF of the Eulerian velocity magnitude and the PDF of the s-Lagrangian
velocity is a direct consequence of the fact that the flow field is
divergence-free. Thus, fluid mass conservation is accounted
for in this sense. 
%%%%%%%%%%%%%%%%%%%%%%%%%%%%%%%%%%%%%%%%%%%%%%%
In summary, geometry and distribution of the
$K$-field are imprinted, at least for small values, in
the distribution of Eulerian velocity magnitudes. 
Based on these observations, we make the following simplifying
assumptions. We consider the spatial distribution of the Eulerian
velocity $v_e(\vx)$ rather than $K(\vx)$ as our starting point. We
note that the small values of velocity magnitude and thus conductivity
dominate the asymptotic transport behavior. Thus, this simplification
allows to study the mechanisms of anomalous transport in correlated
porous media, while the early time behavior is in general not
captured. Thus, we now assume that the Eulerian velocity
field $v_e(\vx)$ is organized in bins of variable horizontal and
constant vertical extensions as described above. In the following, we
specify the heterogeneity and correlation scenarios in terms of the
PDF $p_e(v)$ of Eulerian velocities and $p_{\ell}(\ell)$ of horizontal
bin sizes.  
%%%%%%%%%%%%%%%%%%%%%%%%%%%
\subsubsection{Heterogeneity} 
%%%%%%%%%%%%%%%%%%%%%%%%%%%
We consider two different distributions of $v_e$. The weak heterogeneity scenario is defined by the log-normal 
velocity PDF
\begin{align}\label{exp:v}
 p_e(v) = \frac{1}{v \sqrt{2\pi \sigma_e^2}} \exp{\left\{-\frac{[\ln(v) - \mu_e]^2}{2\sigma_e^2}\right\}}.
\end{align}
where $\mu_e$ is the geometric mean of $v_e$ and $\sigma_e$ the variance of $\ln(v_e)$. Note that the point distribution of hydraulic conductivity is often modeled as a log-normal distribution~\cite{Rubin-2003-Applied}. We consider moderate heterogeneity characterized by $\sigma_e^2 = 1$. The corresponding PDF of the s-Lagrangian velocities $v_s$ is obtained from~\eqref{Eul_sLag} by flux-weighting as
\begin{align}\label{exp:v}
 p_s(v) = \frac{1}{v \sqrt{2\pi \sigma_e^2}} \exp{\left\{-\frac{(\ln(v) - \mu_s]^2}{2\sigma_e^2}\right\}},
\end{align}
where $\mu_s = \mu_e + \sigma_e^2$. 

In order to investigate the impact of strong heterogeneity of velocity point values, we consider a velocity distribution that is characterized by power-law behavior at low velocities~\cite{Dentz:2016aa,tyukhova2016}
\begin{align}
p_e(v) \propto \frac{1}{v_0}\left(\frac{v}{v_0}\right)^{\gamma-1},
\end{align}
and a sharp cut-off for $v \gg v_0$. We consider exponents $0 < \gamma < 1$ and also $-1 < \gamma < 0$. In the latter case, it is understood that the Eulerian velocity PDF has another cut-off at low velocity values, otherwise it is not normalizable. The corresponding PDF of s-Lagrangian velocities is again obtain from~\eqref{Eul_sLag} and behaves at small values as
\begin{align}
\label{power:v}
p_s(v) \propto \frac{1}{v_0}\left(\frac{v}{v_0}\right)^{\beta-1},
\end{align}
where $\beta = \gamma +1$ is between $0$ and $2$. Note that no lower cut-off is needed for values of $\beta$ between $0$ and $1$. For the numerical simulations and detailed analytical calculations, we employ a Gamma-distribution of velocities, which is characterized by the same properties at small $v$ as~\eqref{power:v} and an exponential cut-off for $v \gg v_0$. 
%%%%%%%%%%%%%%%%%%%%%%%%%%%
\subsubsection{Correlation} 
%%%%%%%%%%%%%%%%%%%%%%%%%%%
\begin{figure}[h]
\begin{center}
  \includegraphics[scale=0.7]{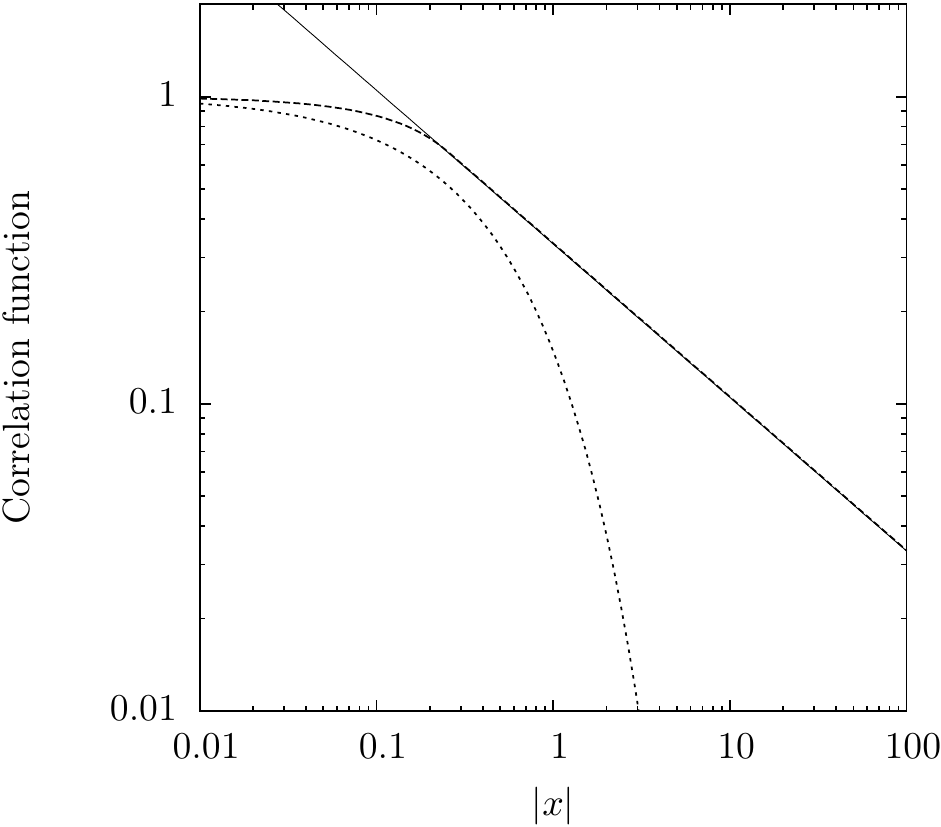} 
\caption{1D correlation function. Comparison between a weakly correlated (dotted line) and a strongly correlated (dashed line) medium. The solid line is $x^{-\frac{1}{2}}$. Results for the case $\ell_0 = \frac{1}{2}$ and 
$\alpha = \frac{1}{2}$, respectively.}
 \label{correlationfunction}
\end{center}
\end{figure}
%%%%%%%%%%%%%%%%%%%%%%%%%%%
The covariance function of the velocity fluctuations $v'_e(\vx) = v_e(\vx) - \langle v_e \rangle$ is defined by 
\begin{align}\label{covfuncdef}
\mathcal C(\vx - \vx') = \langle v'_e(\vx)v'_e(\vx') \rangle. 
\end{align}
The velocity variance is $\sigma_v^2 = \mathcal C(\mathbf 0)$. The correlation function is defined by 
$\mathscr{C}(\vx) = \mathcal C(\vx)/\sigma_v^2$. For the disorder scenarios under consideration it factorizes into 
\begin{align}
\mathscr{C}(\vx) = \mathscr{X}(x) \mathscr{Y}(y) \mathscr{Z}(z),
\end{align}
where $\mathscr{X}(x)$ denotes the correlation function in $x$-direction and $\mathscr{Y}(y)$ and $\mathscr{Z}(z),$ the correlation function in $y$ and $z$-directions, see Appendix~\ref{app:corrfun}. The constant bin size $d_0$ in $y$-direction gives rise to the linear correlation function 
\begin{align}
\label{1Dcorrunif}
\mathscr{Y}(y)=  \left(1-\frac{|y|}{d_0}\right)H(d_0-|y|).  
\end{align}
The same holds for the $z$-direction.
For a general distribution $p_\ell(\ell)$ of bin lengths, we obtain for the correlation function in $x$-direction
\begin{align}
\mathscr{X}(x) = \int\limits_{|x|}^\infty d\ell p_\ell(\ell) \left(1 - \frac{|x|}{\ell} \right), \label{covfunc}
\end{align}
as detailed in Appendix~\ref{app:corrfun}. 

The weakly-correlated scenario is characterized by an exponential distribution of bins sizes
\begin{align}
 p_\ell(\ell) = \frac{\euler^{-\ell/\ell_0}}{\ell_0},
\label{p_ell}
\end{align}
with $\ell_0$ a characteristic scale. 
The correlation function in $x$-direction is then obtained from~\eqref{covfunc} as
\begin{equation}
\mathscr{X}(x) = \euler^{-|x|/\ell_0} + \frac{|x|}{\ell_0}\text{E}_1(-|x|/\ell_0),
 \label{1Dcorrexp}
\end{equation}
where $\text{E}_1(\cdot)$ denotes the exponential integral~\cite{AS1972}. Note that the correlation function decays exponentially at large distance, as shown in Fig. \ref{correlationfunction}. 

The strongly correlated scenario is characterized by a Pareto distribution of bin sizes
\begin{align} \label{power_ell}
 p_\ell(\ell) = \frac{\alpha}{\ell_0} \left(\frac{\ell}{\ell_0}\right)^{-1-\alpha} 
\end{align}
for $\ell > \ell_0$. We consider $0 < \alpha < 2$. Thus, we obtain from~\eqref{covfunc} the correlation function
\begin{equation} \label{1Dcorrpar}
\mathscr{X}(x) = 
\begin{cases}
 \left(\frac{|x|}{\ell_0}\right)^{-\alpha} \left(1-\frac{\alpha}{\alpha+1} \right) & |x| \ge \ell_0 \\
 1 - \frac{\alpha|x|}{\ell_0(\alpha+1)} & |x| < \ell_0.
\end{cases}
\end{equation}
It decays slowly as a power-law for $\ell \geq \ell_0$ as shown in Fig. \ref{correlationfunction}. 
%%%%%%%%%%%%%%%%%%%%%%%%%%
\subsubsection{Ergodicity\label{ergodicity}}
%%%%%%%%%%%%%%%%%%%%%%%%%%
We shortly discuss here the ergodicity of the media model under consideration, this means the equivalence of spatial and 
ensemble sampling of the velocity point values. It is clear that sampling along the $y$-direction is equivalent to ensemble sampling by construction of the random medium. Also, it is clear that spatial sampling along $x$ is equivalent to ensemble sampling for distributions $p_\ell(\ell)$ for which $\langle \ell \rangle < \infty$. Here we briefly discuss the case of $\langle \ell \rangle = \infty$, which is the case for $0 < \alpha < 1$ in~\eqref{power_ell}. The velocity PDF $\hat p_e(v)$ is defined through spatial sampling along the $x$-direction as
\begin{align}
\hat p_e(v) = \lim_{L\to\infty} \frac{1}{L} \int\limits_{-L/2}^{L/2} dx \delta[v-v_e(x)]. 
\end{align}
Because of the geometry of the medium, it can be written as 
\begin{align}\label{alongstream}
\hat p_s(v) =  \lim_{L\to\infty} \frac{1}{L} \sum\limits_{i =0}^{n_L} \ell_i \delta(v-v_i),
\end{align}
where $n_L$ is the number of bins needed to cover the distance $L$. It is given by 
\begin{align}
n_L = \max(n | x_n \leq L), && x_n = \sum\limits_{i=0}^n \ell_i. 
\end{align}
For $0 < \alpha < 1$, the average bin size out of a sample of $n$ scales as $\langle \ell \rangle_n \propto n^{1/\alpha - 1}$, while the average number of bins to cover the distance $L$ is $\langle n_L \rangle \propto L^{\alpha}$~\cite{BG1990}. Thus, we obtain
\begin{align}
\hat p_s(v) =  \lim_{n \to \infty} \frac{1}{n} \sum\limits_{i =0}^{n} \delta(v-v_i) = p_e(v),
\end{align}
this means spatial and ensemble sampling are equivalent. 
%%%%%%%%%%%%%%%%%%%%%%%%%%%%%%%%%%%%%%%%
\section{Average particle motion} \label{coarsegraining}
%%%%%%%%%%%%%%%%%%%%%%%%%%%%%%%%%%%%%%%%
We derive the average particle dynamics based on the streamwise formulation~\eqref{ades} of particle motion. To this end, we disregard particle displacements perpendicular to the mean flow direction, which implies that ${\vv_s(s)}/{v_s(s)}$ is aligned with the $x$-direction. This is justified because the streamline tortuosity is small due to the medium geometry and flow boundary conditions as discussed in Sect.~\ref{disorder}. Furthermore, it has been demonstrated that transverse dispersion is asymptotically zero for purely advective transport in $d = 2$ dimensional porous media~\cite{Attinger:et:al:2004}. We use the geometric structure of the Eulerian velocity to coarse grain the particle motion in time and space. Flow velocities in different bins here are statistically independent. Thus, we coarse grain the distance $s$ along streamlines using the longitudinal bin size as
\begin{align}
s_n = \sum\limits_{i=1}^n \ell_i. 
\end{align}
Thus, we obtain for the space-time particle motion the recursion relations
\begin{align}
\label{x_n+1} 
x_{n+1}  = x_n + \ell_n, 
&&
t_{n+1} = t_n +\frac{\ell_n}{v_n},
\end{align}
where we defined $x_n = x(s_n)$, $t_n = t(s_n)$ and $v_n = v_s(s_n)$. The transition time is defined  by $\tau_n = \ell_n/v_n$. 
We consider a flux weighted extended particle injection at $x = 0$ whose extension is much 
larger than the bin size perpendicular to the flow direction. Thus, the PDF of particle velocities is given by $p_s(v)$ at all steps. The impact of different initial conditions is discussed in~\cite{Dentz2016}. 

The relations~\eqref{x_n+1} define a coupled CTRW~\cite{SL73.1}. Transition time and length are kinematically coupled through velocity, which itself is distributed~\cite{DHSB2008,DentzBolster2010}. This type of coupled CTRW is similar to L\'evy walks~\cite{ShlesingerTurbulence1987,Klafter1990,Meerschaert2009,Rebenshtok2014,zaburdaev2015,DLBLdB:PRE2015} in that transition time and length are kinematically coupled. The L\'evy walk, however, prescribes a transition time PDF $\psi(t)$ and determines the transition length for a constant or distributed velocity kinematically~\cite{zaburdaev2015}. Here, the distribution of transition lengths is dictated by the medium geometry, and the
distribution of velocities by the medium heterogeneity and flow equation as discussed in Sect.~\ref{disorder}. Thus, here the joint PDF $\psi(x,t)$ of transition lengths and times is given in terms of the PDF of transition length and velocities as 
\begin{align}
\psi(x,t) = \int\limits_0^\infty d v \psi(t|x,v)p_\ell(x)p_s(v),
\end{align}
where the conditional PDF of transition time given the transition length and velocity is $\psi(t|x,v) = \delta(t - x/v)$. Evaluating 
the integral gives for $\psi(x,t)$ the expression 
\begin{align}
\psi(x,t) = \frac{x}{t^2} p_\ell(x) p_s\left(\frac{x}{t}\right) \label{psifinal}. 
\end{align}
The marginal PDF of transition times is denoted by $\psi(t)$. The coarse-grained particle position at a given time $t$ is $x_{n_t}$ where $n_t = \sup(n|t_n \leq t)$. Its PDF is given by $P(x,t) = \langle \delta(x - x_{n_t})  \rangle$ where the angular brackets denote the average over all particles in a single realization and the average over the disorder realizations. The evolution of $P(x,t)$ is determined by the following set of equations~\cite{SL73.1,Berkowitz2006} 
\begin{subequations}
\label{cctrw}
\begin{align}
\label{P}
P(x,t) &= \int\limits_0^t dt' R(x,t') \int\limits_{t - t'}^\infty d t'' \psi(t'')
\\
\label{theR}
R(x,t) &= \delta(x) \delta(t) + 
\nonumber\\
& \int dx^\prime \int\limits_0^\infty dt^\prime R(x^\prime,t^\prime)\psi(x-x^\prime, t-t^\prime),
\end{align}
\end{subequations}
where $R(x,t)$ is the probability per time that a particle arrives at a turning point at $(x,t)$. Thus, the right side 
of Eq.~\eqref{P} denotes the probability that a particle just arrives at $x$ at time $t'$ times the probability that the next transition takes longer than $t - t'$. Equation~\eqref{theR} is an expression of particle conservation in $(x,t)$-space.  

Note that $x_{n_t}$ denotes the coarse grained particle position at a turning point of the CTRW. In order to obtain the actual particle position at time $t$, we interpolate by the velocity in the bin such that~\cite{DHSB2008,DentzBolster2010} 
\begin{align}
\label{xt}
x(t) = x_{n_t} + v_{n_t} (t - t_{n_t}), 
\end{align}
where $t_{n_t}$ is the arrival time at the turning point right before $t$. The average particle density is now given by
\begin{equation}\label{c_def}
c(x,t) = \left\langle \delta[x-x_{n_t} - v_{n_t} (t - t_{n_t})]\right\rangle.
\end{equation}
This expression can be expanded to 
\begin{align}
\label{thec}
c(x,t) = \int\limits_0^t dt^\prime \int dx^\prime R(x^\prime,t^\prime) \Phi(x-x^\prime, t-t^\prime),
\end{align}
where $\Phi(x,t) dx$ is the joint probability that the particle makes an advective displacement of a length in $[x,x+dx]$ during time $t$ and that $t$ is smaller than the time for a transition
\begin{equation}
\label{thePhi}
 \Phi(x,t) = \left\langle\delta\left[x -  v_s t\right] \mathbb{I}\left(0\leq t < \ell/v_s \right)\right\rangle .
\end{equation}
The average can be executed explicitly by noting that $\tau = \ell/v_s$ and using the joint PDF $\psi(x,t)$ of transition length and time. This gives 
\begin{align}\label{Phifinal}
 \Phi(x,t) = \int\limits_t^\infty d\tau \frac{\tau}{t}  \psi\left( \frac{\tau}{t}x , \tau\right).
\end{align}

The system~\eqref{cctrw} can be combined into the generalized Master equation for $P(x,t)$ \cite{Berkowitz2002a,KlafterSilbey80}
\begin{align}
\frac{\partial P(x,t)}{\partial t} &= \int dx^\prime \int\limits_0^t dt^\prime \mathcal K(x- x^\prime, t-t^\prime) \nonumber\\ 
& \times[P(x^\prime,t^\prime) - P(x,t^\prime)],\label{sortofmaster}
\end{align} 
where the memory kernel $\mathcal K(x,t)$ is defined through its Laplace transform~\cite{AS1972}
\begin{align}
\mathcal K^\ast(x,\lambda) = \frac{\lambda\psi^\ast(x,\lambda)}{1-\psi^\ast(\lambda)} \, .
\end{align}  
Laplace transformed quantities are marked by an asterisk in the following, the Laplace variable is denoted by $\lambda$. 
We solve for $P(x,t)$ and the particle density $c(x,t)$ in Fourier-Laplace space. We employ here the following definition of the Fourier transform,
\begin{align}
\tilde{c}(k,t) &=  \int  dx \; c(x,t) \exp(i k x), 
\\
c(x,t) &= \int \frac{dk}{2 \pi} \tilde c(k,t) \exp(- i kx). 
\end{align}
Fourier transformed quantities are marked by a tilde, the wave number is denoted by $k$. Thus, we obtain from~\eqref{cctrw} for $\tilde P^\ast(k,\lambda)$
\begin{align}\label{Pstar}
\tilde{P}^\ast(k,\lambda) = \frac{1}{\lambda}\frac{1-\psi^*(\lambda)}{1-\tilde\psi^\ast(k,\lambda)}.
\end{align}
Combining~\eqref{cctrw} and~\eqref{thec} gives for $\tilde c^\ast(k,\lambda)$ 
\begin{equation}\label{cstar}
\tilde{c}^\ast(k,\lambda) = \frac{\lambda\tilde\Phi^\ast(k,\lambda)\tilde{P}^\ast(k,\lambda)}{1-\psi^\ast(\lambda)} \, .
\end{equation}
Equations~\eqref{Pstar} and~\eqref{cstar} form the basis for the derivation of the behaviors of the mean and variance of the particle displacements. 
%
%%%%%%%%%%%%%%%%%%%%%%%%%%%%%%%%%%%%%%
\subsection{Spatial moments}
%%%%%%%%%%%%%%%%%%%%%%%%%%%%%%%%%%%%%%
We study the first and the second centered moment of the particle density $c(x,t)$. While the first moment describes the position of the center of mass, the second centered moment provides a measure of the particle dispersion. Moreover, the temporal scaling of the mean squared displacement is commonly used to discriminate the nature of transport, with non-linear growth being considered a signature of non-Fickian transport. The $j$th moment of $x(t)$ is given by 
\begin{align}
m_j(t) =  \langle x(t)^j \rangle = \int  dx \, x^j c(x,t).
\end{align}
The second centered moment, or in other words, the variance of $x(t)$ is defined by 
\begin{align}
\kappa(t) = m_2(t) - m_1(t)^2. 
\end{align}
In order to calculate the moments, we make use of the following identity in Fourier-Laplace space \cite{Shlesinger1982}
\begin{align}
m^\ast_j (\lambda) &= \left. (-i)^j\frac{\partial^j\tilde{c}^\ast(k,\lambda)}{\partial k^j}\right\rvert_{k=0}.
\label{moments1} 
\end{align}
By substituting \eqref{cstar} into \eqref{moments1} we can express the moments of the particle density $c(x,t)$ in terms of the spatial moments of  $P(x,t)$ and $\Phi(x,t)$. In Appendix \ref{app:momentsinter} we derive the following Laplace space expressions for the first and second displacement moments
\begin{align}
&m_1^\ast (\lambda) = \int\limits_0^\lambda d\lambda^\prime \frac{\mu^\ast_{1} (\lambda^\prime)}{\lambda^2[1-\psi^\ast(\lambda)]}  \label{m1hat} \\
&m_2^\ast (\lambda)=  \int\limits_0^\lambda d\lambda^\prime \frac{2\lambda^\prime\mu_2^\ast(\lambda^\prime)}{\lambda^3[1-\psi^\ast(\lambda)]} +
  \frac{2\mu_1^\ast(\lambda)m_1^\ast (\lambda)}{1-\psi^\ast(\lambda)}, \label{m2hat}
\end{align}
where the $i$th spatial moment of $\psi(x,t)$ is denoted by 
\begin{align}
\mu_i(t) = \int dx \, x^i \psi(x,t). 
\end{align}
%
%%%%%%%%%%%%%%%%%%%%%%%%%%%%%%%%%%%%%%%%
\subsection{First arrival time distribution}
%%%%%%%%%%%%%%%%%%%%%%%%%%%%%%%%%%%%%%%%
The time of first arrival of a particle at a position $x$ is defined by 
\begin{equation}
t_a(x) = \sup{[t| x(t) \leq  x]},
\end{equation}
where $x(t)$ is given by~\eqref{xt}. The arrival time PDF is defined by
\begin{equation}\label{FPTDdef}
f(t,x) = \langle \delta[t-t_a(x)] \rangle .
\end{equation}
Using~\eqref{x_n+1} and~\eqref{xt}, the arrival time can be written as
\begin{align}
t_a(x) = \sum\limits_{i = 0}^{n_x-1} \tau_i + \frac{x - x_{n_x}}{v_{n_x}},
\end{align}
where $x_n$ is given by~\eqref{x_n+1} and $n_x = \sup(n|x_n \leq x)$. The first arrival time PDF satisfies a similar equation as $c(x,t)$ and is given by 
\begin{align}
f(t,x) = &\int\limits_0^x dx^\prime \int\limits_0^t dt^\prime  R(x^\prime, t^\prime) \Theta(x-x^\prime, t-t^\prime),
\end{align}
where $\Theta(x,t)dt$  is the joint probability that the particle makes an advective displacement of length $x$
in a time in the interval $[t,t+dt]$ and that $x$ is smaller than a transition length
\begin{equation}
\Theta(x,t) = \left\langle \delta\left( t-\frac{x}{v}\right) \mathbb{I}(0\leq x < \ell) \right\rangle.
\end{equation}
In analogy with $\Phi(x,t)$, we can relate this joint probability to the joint PDF of
transition lengths and times without interpolation as follows
\begin{equation}
\Theta(x,t)  = x^{-1} \int\limits_x^\infty d\ell \,\ell \psi\left( \ell, \frac{\ell t}{x} \right).
\end{equation}

%%%%%%%%%%%%%%%%%%%%%%%%%%%
\section{Transport behavior} \label{section:transpbehaviour}
%%%%%%%%%%%%%%%%%%%%%%%%%%%
In the following, we study particle dynamics in terms of the variance of particle displacements and first arrival time distributions. 
In order to probe the impact of heterogeneity and spatial correlation on large scale transport, we study three scenarios. 
The first one is characterized by strong heterogeneity and weak correlation, the second by strong correlation and weak disorder. Although driven by different causes, transport in both scenarios is non-Fickian and it exhibits similar behaviors.
The third scenario is characterized by both strong heterogeneity and strong correlation. For each scenario, the transport behavior is investigated through numerical random walk particle tracking simulations of the coarse-grained equations of motion~\eqref{x_n+1}, 
and analytical expressions for the scalings of the moments and the first arrival time distributions.  
%%%%%%%%%%%%%%%%%%%%%%%%%%%%%%%%%%%%
\subsection{Distribution-induced anomalous diffusion}\label{DIAD}
%%%%%%%%%%%%%%%%%%%%%%%%%%%%%%%%%%%%
We first consider the case of anomalous diffusion induced by a broad distribution of velocity point values characterized by the power-law distribution \eqref{power:v}, $p_s(v) \propto v^{\beta - 1}$ for $0< \beta < 2$, and short-range correlation characterized by the exponential distribution of  transition lengths \eqref{p_ell}. This scenario accounts for frequent changes in the particle velocities along trajectories, characterized by the characteristic correlation scale $\ell_0$. 
%%%%%%%%%%%%%%%%%%%%%%%%%%%%%%%%%%%%%%%%
\subsubsection{Dispersion behavior}
%%%%%%%%%%%%%%%%%%%%%%%%%%%%%%%%%%%%%%%%
\begin{figure}[h]
\begin{center}
\vspace{0cm}
\includegraphics[scale = 0.7]{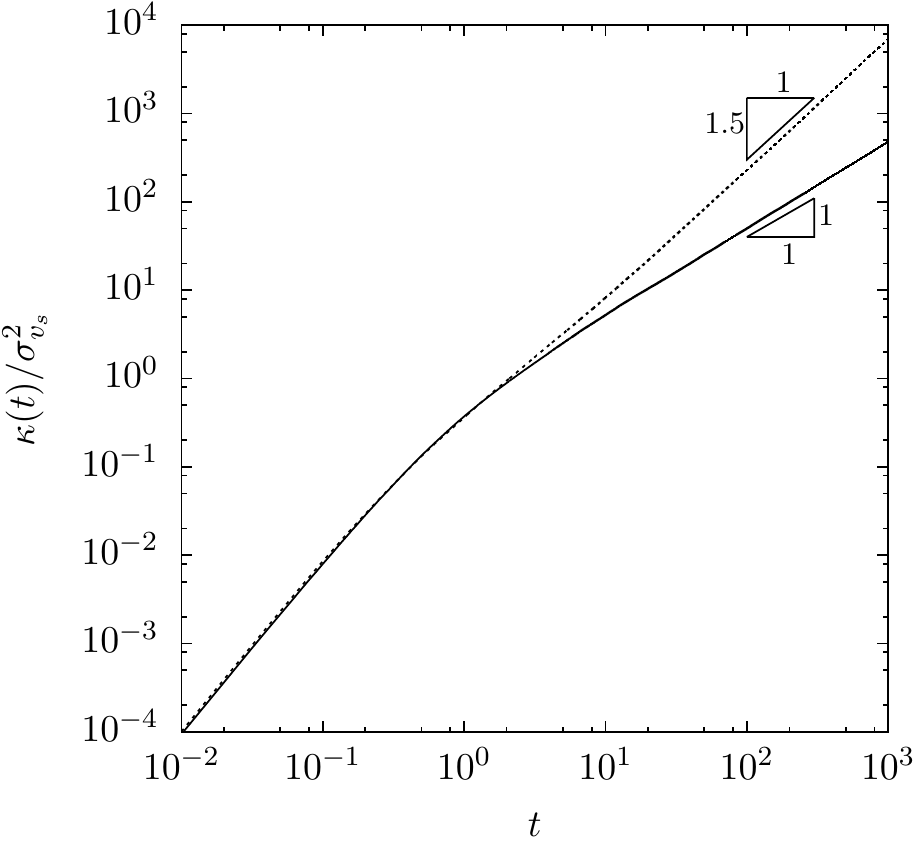}
\caption{Temporal evolution of the variance in case of distribution-induced anomalous transport for $\beta = \frac{1}{2}$ (solid line) and $\beta = \frac{3}{2}$ (dashed line). The second centered moments are normalized by the variance of $v_s$.
} \label{VarExpBin}
\end{center}
\end{figure}
%%%%%%%%%%%%%%%%%%%%%%%%%%%%%%%%%%%%%%%%
The temporal evolution of the mean squared displacement is shown in Fig. \ref{VarExpBin} for two different values of $\beta$ that correspond to different degrees of heterogeneity.  At short times particles move, on average, within a correlation length, where they maintain a constant velocity. As a result, the mean squared 
displacement exhibits a ballistic growth as $\kappa(t) = \sigma_{v_s}^2 t^2$ with $\sigma_{v_s}^2$ the 
variance of the s-Lagrangian velocity $v_s$. The sub ballistic asymptotic behavior depends on the velocity and thus transition time distribution. It arises when the particles have traveled several correlation lengths, thus exploring the heterogeneity of the spatially variable velocity. We observe the same behavior as for an uncoupled CTRW in line with~\cite{DHSB2008}. The explicit expressions for the mean and variance of the particle displacement are derived in Appendix \ref{app:momentscalc1}. For $0 < \beta < 1$, we find that
 \begin{align}
m_1(t) \propto t^\beta & & \kappa(t) \propto t^{2\beta}.
\end{align}
The behavior for $\kappa(t)$ is illustrated in Fig. \ref{VarExpBin} for $\beta = 1/2$. Note that, as discussed in Sect.~\ref{disorder}, this behavior has to be understood in a preasymptotic sense because the Eulerian velocity PDF $p_e(v)$ needs a cut-off at low velocities to be normalizable. 
%%%%%%%%%%%%%%%%%%%%%%%%%
\begin{figure}[h]
\vspace{-0cm}
% \hbox{\hspace{-0cm}\includegraphics[scale=0.60]{integ_expbin-crop.pdf}}
\hbox{\hspace{-0cm}\includegraphics[scale=0.60]{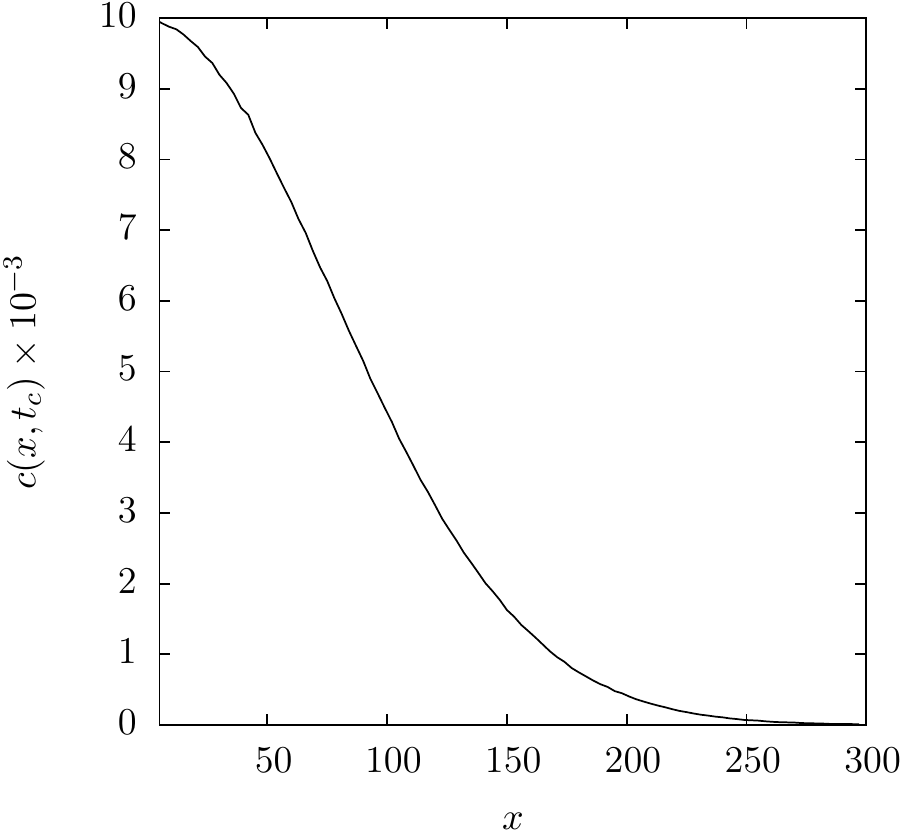}}  \hbox{\hspace{-0cm}\includegraphics[scale=0.60]{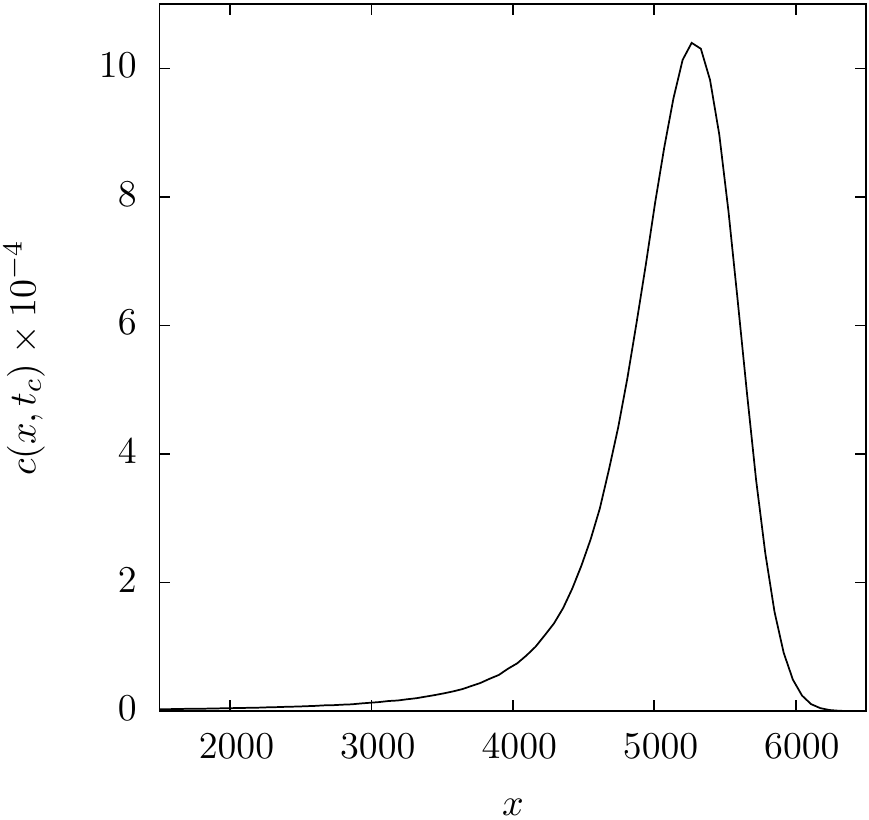}}
\caption{Particle density at time $t_c = 10^4$ for $\beta = \frac{1}{2}$ (upper panel) and $\beta = \frac{3}{2}$ (lower panel). The results are obtained by a CTRW simulation with $10^7$ particles . 
Injection occurs point-wise and impulsively at $x=0$ and $t=0$.} \label{plumeExpbin1.5}
\end{figure}
%%%%%%%%%%%%%%%%%%%%%%%%%
For $1 < \beta < 2$, we derive for the displacement mean and variance the scalings
\begin{align}
m_1(t) \propto t & & \kappa(t) \propto t^{3-\beta}.
\end{align}
The behavior for $\kappa(t)$ is illustrated in Fig. \ref{VarExpBin} for $\beta = 3/2$. 
These results are consistent with those for uncoupled CTRW \cite{Shlesinger74,MABE02,DCSB2004}.

Figure~\ref{plumeExpbin1.5} shows the particle distributions for $\beta = 1/2$ and $\beta = 3/2$. Due to the high probability of low velocities, $c(x,t)$ has a forward tail and strong localization at the origin. For $\beta = 3/2$, particles are more mobile, which manifests in a leading front and a trailing tail. 
%%%%%%%%%%%%%%%%%%%%%%%%%%%%%%%%%%%%%%
\subsubsection{First arrival time distribution} 
%%%%%%%%%%%%%%%%%%%%%%%%%%%%%%%%%%%%%%
%%%%%%%%%%%%%%%%%%%%%%%%%%%%%%%%%%%%%%
\begin{figure}[h]
\includegraphics[scale=0.7]{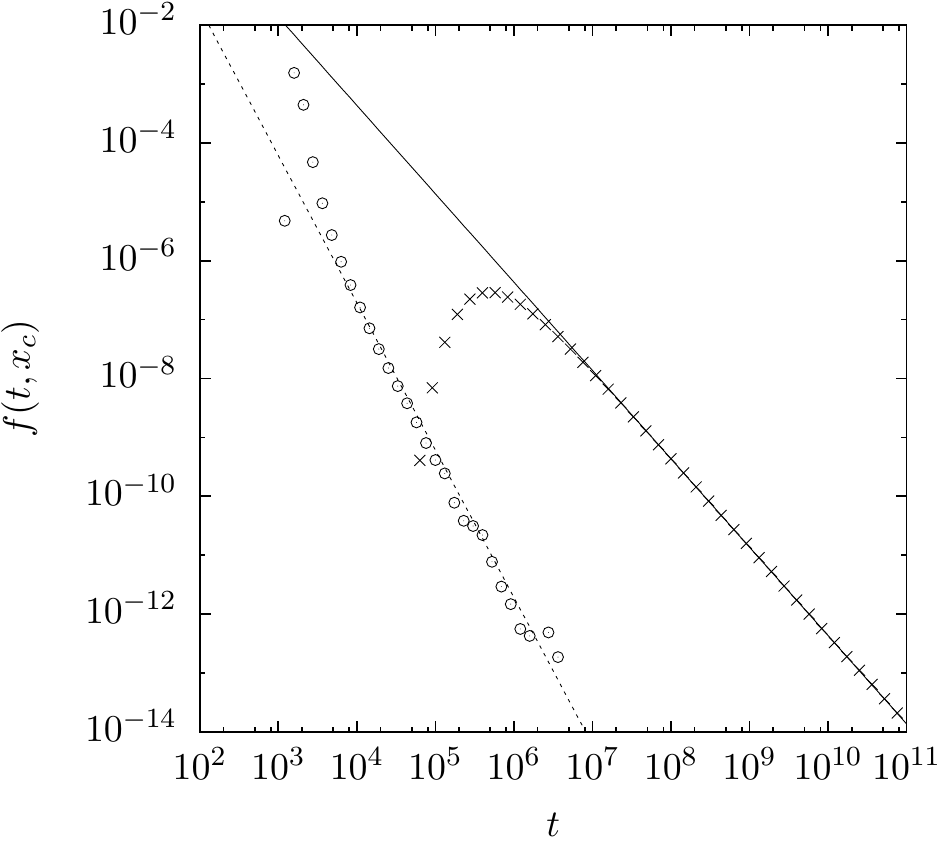}
\caption{First arrival time distribution for the cases $\beta = \frac{1}{2}$ (crosses) and $\beta = \frac{3}{2}$ (circles). The results are obtained by CTRW 
simulations using $10^7$ particles. The solid line is $t^{-\frac{3}{2}}$, while the dashed line is $t^{-\frac{5}{2}}$. Particle injection occurs instantaneously at the origin of space and time and the detection is performed at 
$x_c=100\ell_0$.} \label{FPT_constbin}
\end{figure}
%%%%%%%%%%%%%%%%%%%%%%%%%%%%%%%%%%%%%%
Figure \ref{FPT_constbin} shows the first arrival time distributions for the exponents $\beta = 1/2$ and $\beta = 3/2$ at a distance of $x_c = 10^2 \ell_0$ from the injection point. Again the case $0< \beta < 1$ needs to be understood in a preasymptotic sense. The peak of the arrival time distribution for $\beta = 1/2$ is strongly delayed compared to the one for $\beta = 3/2$ due to the higher probability of low velocities. The tailing behavior is characterized by $f(t,x_c) \propto t^{-1-\beta}$ characteristic for an uncoupled CTRW. This behavior can be readily understood as follows. The average number of steps $n_c$ needed to arrive at the control point is $x_c/\ell_0$. The transition time may be approximated by $\tau \approx \ell_0/v_s$, so that the transition time PDF is approximately
\begin{align}
\psi(t) \approx \frac{\ell_0}{t^2} p_s(\ell_0/t) \propto t^{-1-\beta}
\end{align}
for $t \gg \ell_0 / v_0$. We used \eqref{power:v} for $p_s(v)$. The tailing behavior of $f(t,x_c)$ follows for $0< \beta < 2$ from the generalized central limit theorem. 
%%%%%%%%%%%%%%%%%%%%%%%%%%%%%%%%%%%%%%%%%%%%%%%%%%%%%%%%
\subsection{Correlation-induced anomalous diffusion}\label{CIAD}
%%%%%%%%%%%%%%%%%%%%%%%%%%%%%%%%%%%%%%%%%%%%%%%%%%%%%%%%
Here we study the case of anomalous diffusion induced by correlation. To this end we consider the power-law distribution of transition lengths \eqref{power_ell}, $p_\ell(\ell) \propto \ell^{-1-\alpha}$ for $0 < \alpha < 2$, and the log-normal distribution of velocities \eqref{exp:v} for $\sigma_2^2  =1$ and $\mu_e = 0$. Following the path of the previous section, we study the temporal evolution of the spatial moments and the first arrival time distribution to understand the impact of correlation on the average transport.  
%%%%%%%%%%%%%%%%%%%%%%%%%%%%%%%%
\subsubsection{Dispersion behavior}
%%%%%%%%%%%%%%%%%%%%%%%%%%%%%%%%
\begin{figure}[h]
\begin{center}
\vspace{0cm}
\includegraphics[scale = 0.7]{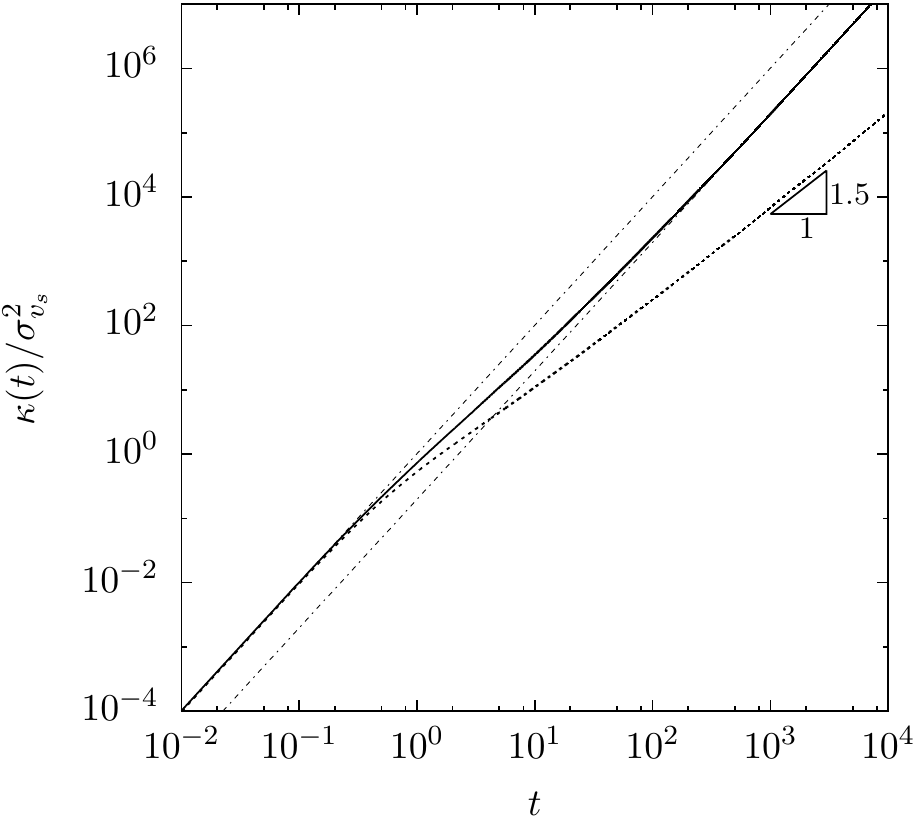}
\caption{Temporal evolution of the variance in case of correlation-induced anomalous transport for $\alpha = \frac{1}{2}$ (solid line) and $\alpha = \frac{3}{2}$ (dashed line). Dashed-dotted lines indicate ballistic growth.
}\label{VarExpTheta}
\end{center}
\end{figure}
Figure \ref{VarExpTheta} shows the temporal evolution of $\kappa(t)$ for two different values of $0< \alpha < 1$ and $1 < \alpha < 2$. The degree of correlation is determined by the exponent $\alpha$. At early times, most of the particles have traveled less than a correlation length and, as a consequence, they have maintained their initial velocity. The early time behavior of $\kappa(t)$ is
ballistic. The asymptotic scaling behaviors are derived in Appendix \ref{app:momentscalc2}. 

\begin{figure}[h]
\vspace{-0cm}
\hbox{\hspace{-0cm}\includegraphics[scale=0.7]{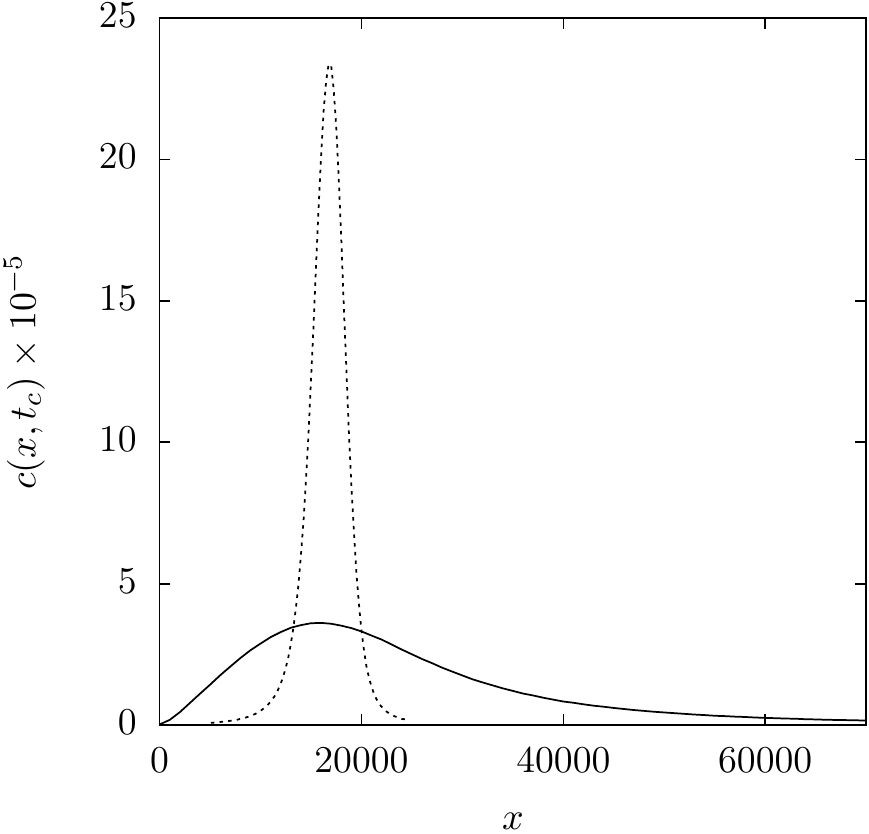}}
\caption{Particle density at time $t_c = 10^4$ for $\alpha = \frac{1}{2}$ (solid line) and $\alpha = \frac{3}{2}$ (dotted line). The results are obtained by CTRW simulations with $10^7$ particles . Injection occurs
point-wise and impulsively at $x = 0$ and $t = 0$.
 } \label{plumeExptheta}
\end{figure}

For very strong correlation, this means $0 < \alpha < 1$, we obtain 
\begin{align}
  m_1(t) \propto t & & \kappa(t) \propto t^2. 
\end{align}
While the center of mass position increases linearly with time, the variance shows still ballistic behavior. This is a consequence of the broad distribution of correlation scales. While a given proportion of particles have changed velocities at asymptotically long time, a large proportion still persists in the initial velocity. In fact, for $0 < \alpha < 1$, the mean transition length is infinite and the number of velocity changes increase sublinearly with distance $x$ as $x^{\alpha}$, see also the discussion in Sect.~\ref{ergodicity}. The number of velocity changes corresponds to the number of bins needed to cover the distance $x$. The resulting ballistic behavior of the persistent particles dominates over the dispersion of the particle that have experienced several velocity transitions. The spatial particle distribution for $\alpha = 1/2$ is shown in Fig. \ref{plumeExptheta}. Initial difference in the particle velocities are amplified with time due to their persistence. The spatial distribution reflects the distribution of velocities $p_s(v)$. 

For values of $\alpha$ between $1$ and $2$, correlation is still strong, but here the mean transition length is finite.  
We obtain the following scalings for the mean and variance of the particle displacements
\begin{align}
  m_1(t) \propto t & & \kappa(t) \propto t^{3-\alpha}.
\end{align}
Because of the strong correlation, those particles that experience low velocities as they move through regions of low conductivity are efficiently separated from those that move fast. Although the heterogeneity is weak and the velocities show small variability, those velocities are  kept for a long distance. The resulting separation of particles gives rise to the superdiffusive behavior. The corresponding particle density for $\alpha = 3/2$ is shown in Fig. \ref{plumeExptheta}. Unlike for disorder dominated superdiffusion, see Fig.~\ref{plumeExpbin1.5}, here the particle distribution does not show a dominant backward tail. Superdiffusion is due to persistent velocity contrast and not to slow velocities.  
%%%%%%%%%%%%%%%%%%%%%%%%%%%%%%%%%%%%%%%%%%%%%
%%%%%%%%%%%%%%%%%%%%%%%%%%%%%%%%%%%%%%%%%%%%%

\subsubsection{First arrival time distribution}\label{sectionFPTDweakHet}
\begin{figure}
\includegraphics[scale = 0.7]{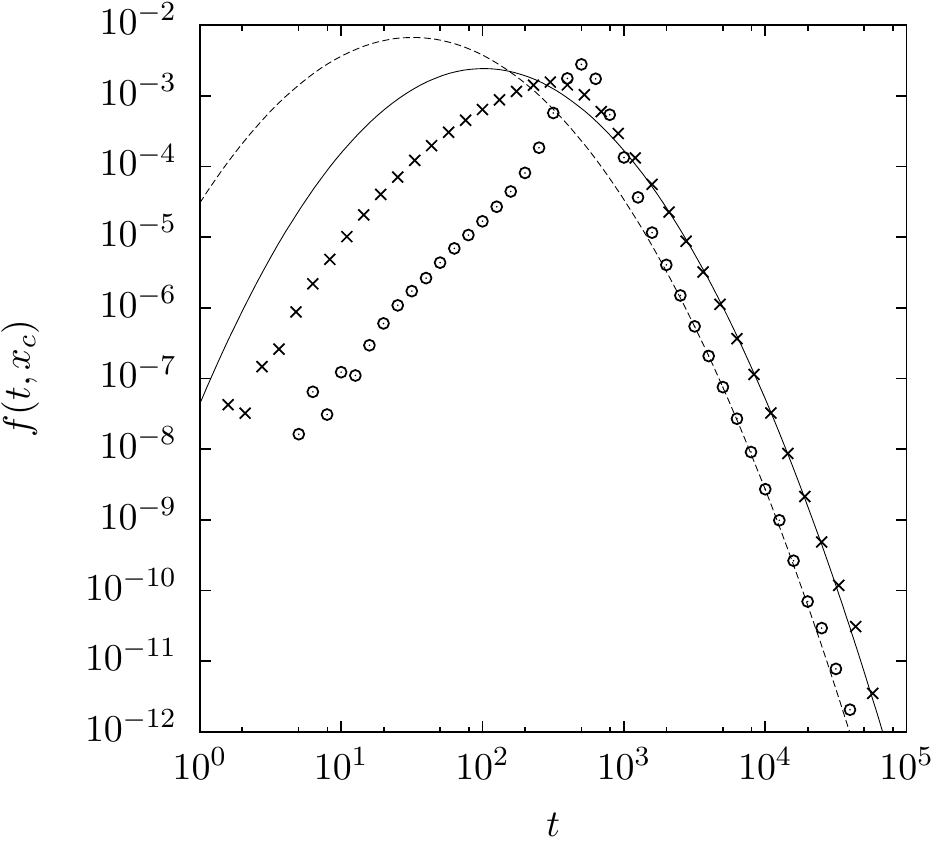}
\caption{First arrival time distribution for the cases $\alpha = \frac{1}{2}$ (crosses) and $\alpha = \frac{3}{2}$ (circles). The solid and the dashed lines are log-normal distributions fitted to the data. The results are obtained
by CTRW simulations using $10^8$ particles.  injection occurs instantaneously at the origin of space and time and the detection is performed at $x_c=100\ell_0$.} \label{FPTD_corr}
\end{figure}

Figure \ref{FPTD_corr} shows the first arrival time distributions for $\alpha = \frac{1}{2}$ and $\alpha = \frac{3}{2}$ at a detection plane located at a distance $x_c = 10^2\ell_0$ from the inlet. We observe an earlier peak for the case $\alpha = \frac{1}{2}$, which is due to those particles that maintain a high velocity for a long distance, since this case corresponds to the higher correlation. At late times, both curves show log-normal tailings. This kind of behavior is particularly interesting if compared to the results of dispersion. In fact, although the variance exhibits a super-linear growth in time, no anomalous behavior is observed in the first arrival time distribution. In order to explain this character, we recall that, due to the high variability of bins lengths, a significant proportion of particles travels until the detection plane $x_c$ without performing any transition, i.e. by keeping the same initial velocity. This proportion of particles is given by 
\begin{align}\label{proportion}
P_0(x_c) = \int_{x_c}^\infty d\ell \, p_\ell(\ell). 
\end{align}
For the distribution of Eq. \eqref{power_ell} we obtain $P_0(x_c) = \left(\frac{\ell_0}{x_c}\right)^\alpha$.
For these particles, the arrival time at $x_c$ is given by the kinematic relationship $t_a = x_c/v$. Thus, the first arrival time distribution can be written as
\begin{align} \label{fatd}
f(t,x_c) = P_0(x_c) \frac{x_c}{t^2}p_s\left(\frac{x_c}{t} \right) + ...,
\end{align}
where the dots indicate the contribution by particles undergoing transitions. 
Since the distribution of velocities is log-normal, $f(t,x_c)$ is asymptotically also log-normal and this explains the tails that we observe in Fig. \ref{FPTD_corr}.
Because for $\alpha = \frac{1}{2}$ the proportion of particles that undergo no transitions is larger than for $\alpha = \frac{3}{2}$, the log-normal tailing arises earlier.
\subsection{Anomalous diffusion induced by distribution and correlation}
In this last scenario, we study anomalous diffusion induced by both distribution and correlation. In order to do so, we consider distributions with power-law tails for both the transition lengths \eqref{power_ell} for $0 < \alpha < 2$ and the velocities \eqref{power:v} for $0<\beta<2$. As we discussed, the case $0<\beta<1$ has to be intended in a preasymptotic sense.
 As we did in the previous sections, we analyze the behavior of the spatial moments and the first arrival time distribution to quantify the impact of strong correlation and strong distribution on transport.
\subsubsection{Dispersion behavior} \label{strongstrongdisp}
\begin{figure}[h]
\begin{center}
\vspace{0cm}
\includegraphics[scale = 0.7]{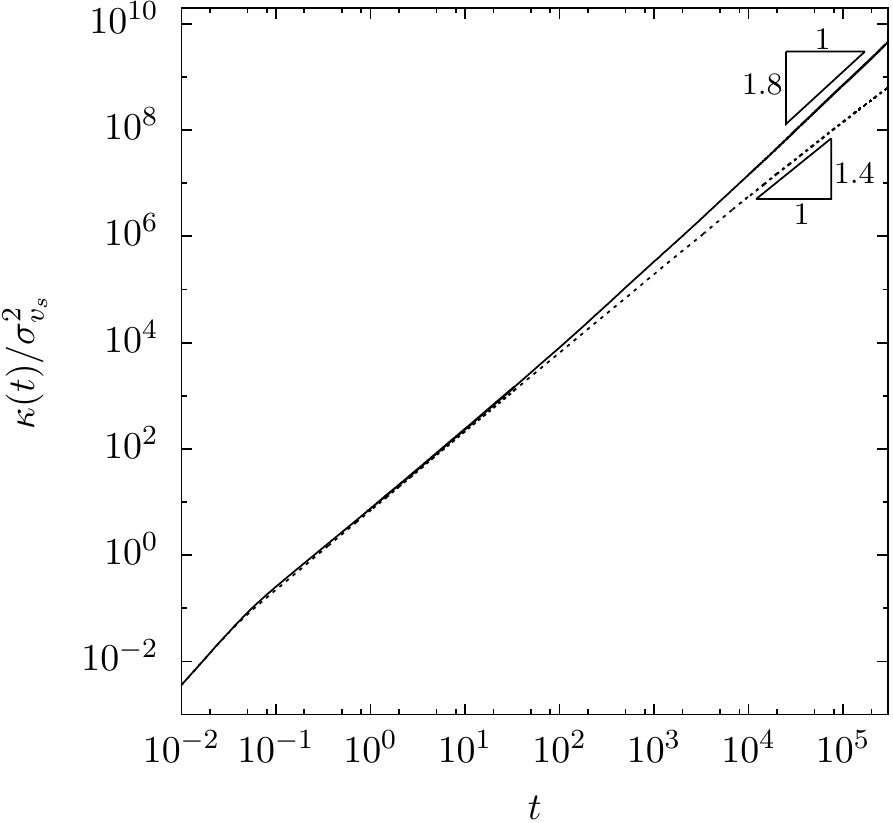}
\caption{Temporal evolution of the variance in case of anomalous transport induced by both distribution and correlation for $\alpha = 1.4, \beta = 1.2$ (solid line) and for
$\alpha = 1.6, \beta = 1.9$ (dashed line).}\label{Var1515}
  \end{center}
\end{figure}

Figure \ref{Var1515} shows the temporal evolution of the mean squared displacement for two different choices of the shape parameters $\alpha \in (1,2)$ and $\beta\in (1,2)$. In particular, the cases $\alpha < \beta$ and $\alpha > \beta$ are considered in order to understand the relative impact of each process. 
At early times, particles have traveled less than a correlation length. Thus, the variance exhibits a ballistic behavior, as particles have maintained their initial velocity. In the large time limit, we observe a convergence to the asymptotic regimes that are derived analytically in Appendix \ref{app:momentscalc}.
We obtain for this case
\begin{align}
m_1(t) \propto t & & \kappa(t) \propto t^{3-\omega},
\end{align}
where $\omega = \min{(\alpha,\beta)}$. This means that the asymptotic behavior is determined by the stronger between disorder and correlation. Thus, for $\alpha < \beta$ the superdiffusive behavior is due to the persistent contrast of velocities, rather than on the retention of particles with slow velocities. Conversely, for $\alpha > \beta$, the impact of slow velocities becomes more important than the persistence of different velocities.
 
\begin{figure}[h]
\begin{center}
\vspace{-0cm}
\hbox{\hspace{0cm}\includegraphics[scale=0.7]{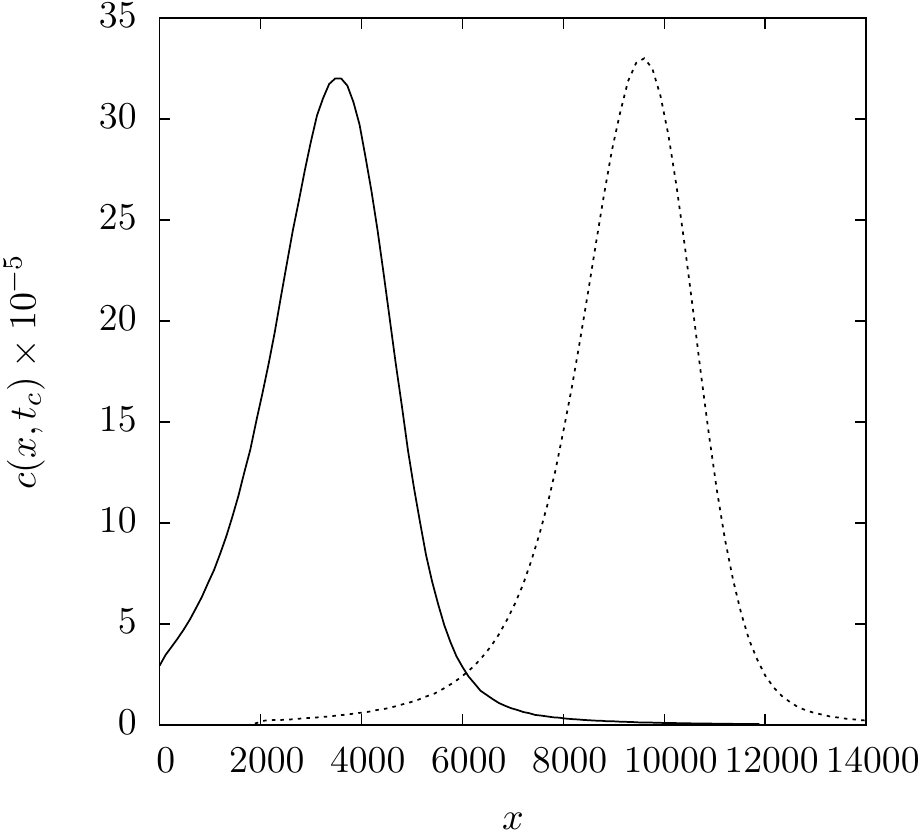}}
\caption{Particle density at time $t_c = 10^4$ for $\alpha = 1.4, \beta = 1.2$ (solid line) and for $\alpha = 1.6, \beta = 1.9$ (dotted line). The results are obtained by
CTRW simulations with $10^7$ particles . Injection occurs point-
wise and impulsively at $x = 0$ and $t = 0$.} \label{plumeab1.5}
\end{center}
\end{figure}

Figure \ref{plumeab1.5} shows the spatial particles density for the two considered cases. We observe that the peak position depends on the value of $\beta$, since smaller values correspond to an higher probability of low velocities and, thus, to a retarded peak. We also observe that the curves are tailed towards the same direction, but the processes that lead to this phenomenon are opposite. For $\alpha < \beta$, correlation is stronger than distribution and the tail develops itself towards low values, in analogy to what we observed in Sect. \ref{CIAD}. For $\alpha > \beta$, distribution dominates over correlation. We observe that the same tailing as for the case of distribution-induced anomalous diffusion (see Fig. \ref{plumeExpbin1.5}, lower panel).

Until here we have considered the case in which both $\alpha$ and $\beta$ are between $1$ and $2$. Nevertheless, a variety of different cases may arise. In the following, we discuss different scenarios related to different choices of the exponents $\alpha$ and $\beta$, which means different degrees of correlation and disorder. The scalings of the moments are derived in Appendix \ref{app:momentscalc}.

\paragraph{Case $\boldsymbol\alpha \boldsymbol\in \mathbf{(0,1)}$,
  $\boldsymbol\beta\boldsymbol\in\mathbf{(1,2)}$} 

In this case, we derive that the first moment and the variance scale as 
\begin{align}
m_1(t) \propto t & & \kappa(t) \propto t^2.
\end{align}
This scenario is tantamount to the case of correlation-induced anomalous diffusion with $0<\alpha<1$ described in Sect. \ref{CIAD}. 
Since no mean transition length exists, transport behavior is fully determined by the longest bins and, consequently, dispersion is
ballistic. The net effect is that the process \eqref{x_n+1} is decoupled.

\paragraph{Case $\boldsymbol\alpha \boldsymbol\in \mathbf{(1,2)}$,
  $\boldsymbol\beta\boldsymbol\in\mathbf{(0,1)}$} 

We derive the following scalings for the first moment and the variance of particles displacement
\begin{align}
m_1(t) \propto t^\beta & & \kappa(t) \propto t^{2\beta}.
\end{align}
Notice that these scalings are the same that we observed in Sect. \ref{DIAD} for $0<\beta<1$. The reason for this fact is that this case is dual to the previous. In fact, while on one hand a mean transition length can be defined, on the other no mean transition time exists. Thus, transport is dominated by disorder and it exhibits a non-Fickian behavior $\kappa(t) \propto t^{2\beta}$ that is due to the retention of particles moving with low velocities. The strength of retention depends on the exponent $\beta$. In particular, we observe subdiffusive behavior for $0< \beta < 1$ and superdiffusive growth for $1<\beta<2$.

\paragraph{Case $\boldsymbol\alpha\mathbf{,}\,\boldsymbol\beta\boldsymbol\in\mathbf{(0,1)}$} In this case, the scalings of the moments depend on the relationship between $\alpha$ and $\beta$. In particular, we get that the mean and the variance of particles displacement scale as
\begin{align}
m_1(t) \propto t^\nu & & \kappa(t) \propto t^\epsilon,
\end{align}
where $\nu = \min(1, \beta - \alpha +1)$ and $\epsilon = \min(2, 2 + \beta -\alpha)$. This means that for $\alpha < \beta$ we get ballistic growth of the variance, which is analogous to the behavior that we observed in Sect. \ref{CIAD} for $0<\alpha<1$. It is interesting to observe that for $\alpha > \beta$, a superballistic behavior arises. This very anomalous behavior is due to the combined action of very low velocities and the high persistence of the velocity contrast. However, we recall that the case $0<\beta< 1$ has to be understood in a preasymptotic sense.
%
% %
\subsubsection{First arrival time distribution}
\begin{figure}[h]
\vspace{-0cm}
\hbox{\hspace{0cm}\includegraphics[scale=0.7]{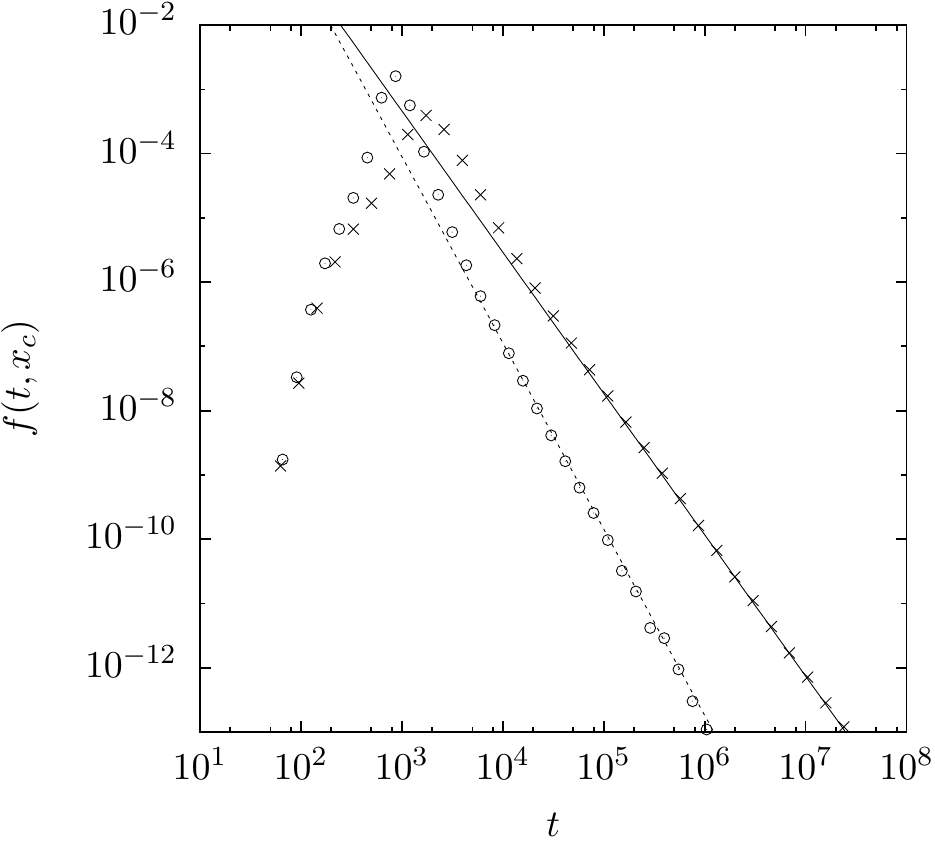}}
\caption{First arrival time distribution for $\alpha= 1.4 ,\beta = 1.2$ (crosses) and for $\alpha = 1.6, \beta = 1.9$ (circles). The results are obtained by CTRW 
simulations with $10^8$ particles. The solid line is $t^{-2.2}$, while the dashed line is $t^{-2.9}$. The injection occurs instantaneously at $x=0$ and $t=0$ and the detection is performed at $x_c=100\ell_0$. } \label{FPT_05}
\end{figure}
%
% %
Figure \ref{FPT_05} shows the first arrival time distribution for two combinations of $\alpha$ and $\beta$ between $1$ and $2$. We distinguish the cases $\alpha < \beta$ and $\alpha > \beta$, as they correspond to the cases in which the dominating processes are correlation and disorder, respectively. The positions of the peaks appear shifted. This is due to the fact that for smaller values of $\beta$ the probability of encountering very low velocities is higher.
In both the considered cases, the first arrival time distribution behaves asymptotically as
\begin{equation}
f(t,x_c) \propto t^{-1-\beta}.
\end{equation}

As we discussed in Sect. \ref{CIAD}, the tails of the distribution are determined by those particles that undergo no velocity transitions. The relative proportion of these particles is given by Eq. \eqref{proportion}. Thus, the first arrival time distribution is given by Eq. \eqref{fatd}. Since the distribution of velocities scales as $v^{\beta-1}$, the distribution of arrival times scales as $t^{-1-\beta}$ due to the kinematic relationship $t_a = x_c/v$.

\section{Summary and conclusions} \label{summary}
We investigate the origins of anomalous transport in the flow through
correlated porous media focusing on the impact of disorder and
correlation. We consider quenched $d$-dimensional random hydraulic
conductivity fields, in which the correlation structure is  determined
by a distribution of length scales of regions of equal hydraulic
conductivity $K$. The spatial variability in $K$ is mapped onto a
distribution of Eulerian velocities through the Darcy
equation. Particle transport is characterized by the series of
Lagrangian velocities sampled equidistantly along the streamlines,
whose statistics are related to the Eulerian velocity PDF by flux-weighting. 
We show that average particle follows a coupled CTRW characterized by
the PDF of characteristic length scales and the PDF of Eulerian
velocities. Within this framework, we derive analytical expressions
for the asymptotic scaling of the moments of particle displacements
and the first arrival time distributions or breakthrough curves. 
In order to quantify the impact of disorder and correlation on average
transport, we consider three different scenarios, in which the
anomalous behaviors are induced by disorder, correlation or both.

In the first scenario, we use an exponential distribution of bin sizes and a Gamma distribution of velocities $p_v(v) \propto v^{\beta-1}$. Since the transition length PDF is sharply peaked, in the long time limit
this case is equivalent to an uncoupled CTRW. Thus, we get that the mean squared displacement evolves in time as $\kappa(t) \propto t^{2\beta}$ for $\beta\in(0,1)$ and $\kappa(t)\propto t^{3-\beta}$ 
for $\beta \in (1,2)$. The first arrival time distribution exhibits retarded peaks for smaller values of $\beta$ that are due to the higher probability of having lower velocities and a tail proportional to $t^{-1-\beta}$ which is a consequence of the generalized central limit theorem.

The second scenario accounts for the effects of strong correlation, which is modeled through a power-law distribution of bin sizes $p_\ell(\ell) \propto \ell^{-1-\alpha}$. For $\alpha \in (0,1)$, because the mean correlation 
length is infinite, transport is dominated by those particles that undergo no transition. This reflects itself into the observation of a ballistic growth of the mean squared displacement and breakthrough curves that behave asymptotically as the distribution of the inverse of velocities. 
For $\alpha \in (1,2)$, because a mean correlation length can be defined, particles undergo velocity transitions in a finite time. Nevertheless, some very long 
bins with low velocities may be encountered, which gives rise to an efficient retention of particles that manifests itself in the stretching of the spatial density distribution and in the superlinear
growth of the mean squared displacement $\kappa(t) \propto t^{3-\alpha}$. In this case, the anomalous character is not determined by low velocities, but by the persistence of velocity contrasts for long distances.

In the last scenario that we consider, power-law distributed bin sizes
and velocities are used. We distinguish a variety of asymptotic
behaviors that depend on the exponents $\alpha$ and $\beta$ and thus
on the relative importance of correlation versus disorder
distribution.  The transport behavior is in general governed by the
process characterized by the heavier tails. For example, for $\alpha$
and $\beta$ between $1$ and $2$, the velocity distribution dominates
for $\beta < \alpha$, while for $\alpha < \beta$ correlation
determines the asymptotic behavior of the displacement variance. The
long time behavior of the particle arrival times is again dominated by
particles with persistent velocities, this means, particles that have
not made a velocity transition until the sampling position. The
arrival time distribution thus scales as the PDF of inverse
velocities.

In conclusion, we have characterized anomalous behaviors of transport
in correlated porous media. These non-Fickian behaviors are induced by
heterogeneity and correlation. We show that in some cases it is not
possible to decouple the effects of these processes, even though in
general the stronger process determines the nature of transport. This
work sheds some new light on th mechanisms underlying anomalous
transport in porous media, which may aid in identifying their
footprints from experimental data. Future work will address
the generalization of the derived approach in presence of  diffusion and local
scale dispersion. 

\paragraph{Acknowledgements} The support of the European Research Council (ERC) through the project MHetScale (617511) is gratefully acknowledged.

\section*{Author contribution statement} 
The authors contributed equally to the paper.
 \appendix
%

%%%%%%%%%%%%%%%%%%%%%%%%%%%%%%%%%%%%%%%%%%%%%%%%%%%%%
\section{Eulerian and s-Lagrangian velocity PDFs}\label{EusLag}
%%%%%%%%%%%%%%%%%%%%%%%%%%%%%%%%%%%%%%%%%%%%%%%%%%%%%
Here we show the derivation of the s-Lagrangian PDF from the Eulerian PDF. The latter is defined as
\begin{align}\label{eulerianPDF}
 p_e(v) = \lim_{V\to\infty} \frac{1}{V} \int_\Omega d\vx \,\delta[v - v_e(\vx)]\,, 
\end{align}
where $V$ is the volume of the region $\Omega$. We define the s-Lagrangian PDF sampled among particles as
\begin{align}\label{sLag_part}
p_s(v,s) = \lim_{V_0\to\infty} \frac{1}{V_0} \int_{\Omega_0} d\va
  \frac{v(\va)}{\langle v_e \rangle} 
\delta\left( v - v_e[\vx(s,\va)]  \right),
\end{align}
where $\vx(s = 0;\va) = \va$, $\Omega_0$ is the region
of space occupied by the particles at $s = 0$, $V_0$ its volume. % , and
% %
% \begin{align}
% A_0 = \frac{1}{V_0} \int_{\Omega_0} d\va v(\va).  
% \end{align}
% %
% Note that $A_0 \to \langle v_e \rangle$ in the limit $V_0 \to
% \infty$.
Expression~\eqref{sLag_part} accounts for the flux-weighting
of the initial particle injection. We apply the transformation $\vx =
\vx(s,\va)$ to Eq. \eqref{eulerianPDF} in order to obtain
\begin{align}\label{sLag}
p_e(v) = &\lim_{V\to\infty} \frac{1}{V} \int_{\Omega_0} d\va \, \mathbb{J}(\va,s) 
\delta\left(v-v_e[\vx(s,\va)]\right)\,,
\end{align}
where $\mathbb{J}(\va,s)$ is the norm of the determinant of the
Jacobian of the transformation. We notice that the following
relationship holds $\frac{d}{ds}\mathbb{J}  =  \mathbb{J} \nabla \cdot
\left(\frac{\vv_s}{v_s}\right)$. Under the condition of
incompressibility, the latter reduces to
\begin{align}\label{sLag_incompress}
\frac{d}{ds}\mathbb{J}  =  -\mathbb{J} \frac{\vv_s\cdot\nabla v_s}{v_s^2}\,.
\end{align}
Since $v_s(s) = v_e[\vx(s,\va)] = \lVert \vu[\vx(s;\va)] \rVert$, we
obtain from Eq.~\eqref{xst}
\begin{align}
 \frac{dv_s}{ds} = \frac{\vv_s}{v_s}\cdot\nabla v_s. 
\end{align}
Thus, Eq. \eqref{sLag_incompress} reduces to
\begin{align}\label{sLag_variable}
\frac{d}{ds}\mathbb{J}  = -\frac{1}{v_s} \frac{dv_s}{ds}\mathbb{J} \,.
\end{align}
Since for $\mathbb{J}(\va,0)=1$, corresponding to the fact that the
starting points are mapped identically onto themselves for $s=0$,
integrating the differential Eq. \eqref{sLag_variable} yields
\begin{equation}
 \mathbb{J}(\va,s) = \frac{v_e(\va)}{v_e[\vx(s,\va)]}. 
\end{equation}
By substituting the latter into Eq. \eqref{sLag}, we obtain  
\begin{align}\label{fwPDF}
 p_e(v) = \lim_{V \to \infty} \frac{1}{V} \int_{\Omega_0} d\va \, v_e(\va)
  \frac{\delta\left(v-v_e[\vx(s,\va)]\right)}{v_e[\vx(s,\va)]}. 
\end{align}
We can write this expression as 
\begin{align}\label{fwPDF:2}
p_e(v) &= 
\nonumber\\
& \frac{\langle v_e \rangle}{v} \lim_{V \to \infty} \frac{1}{V} \int_{\Omega_0}
  d\va \frac{v_e(\va)}{\langle v_e \rangle} \delta\left(v-v_e[\vx(s,\va)]\right), 
\end{align}
where we use that $v = v_e[\vx(s,a)]$ as per the Dirac delta in the
integrand. Using~\eqref{sLag_part} to identify
$p_s(v)$ on the right side gives Eq. \eqref{Eul_sLag}.
%%%%%%%%%%%%%%%%%%%%%%%%%%%%%%%%%%%%%%%%%%%%%%%%%%%%%
\section{Correlation functions}\label{app:corrfun}
%%%%%%%%%%%%%%%%%%%%%%%%%%%%%%%%%%%%%%%%%%%%%%%%%%%%%
In this Appendix, we derive the analytical expressions of the correlation function for the geometry described in Sect. \ref{disorder}. To this scope, we introduce the fluctuations of the Eulerian velocity
with respect to its average $v_e^\prime(\mathbf{x}) = v_e(\mathbf{x}) -\langle{v_e}(\mathbf{x})\rangle$, where the mean of $v_e^\prime$ is null by definition. For a position in the bin $(n,m,p)$, we set $x = x_n +\delta_x$,
$y = y_n +\delta_y$ and $z = z_n + \delta_z$,  where $\delta_x$ is uniformly distributed between $0$ and $\ell_{n+1}$, while $\delta_y$ and $\delta_z$ are uniformly distributed in $(0, d_0]$ and in $(0, h_0]$, respectively. 
Therefore, the fluctuations can be expressed as
\begin{align}
v_e^\prime(\mathbf{x}) = \sum_{n,m,p}&v_{e;n,m,p}^\prime \mathbb{I}(0 < \delta_x \leq \ell_{n+1}) \nonumber \\
&\times  \mathbb{I}(0 < \delta_y\leq d_0)\mathbb{I}(0 < \delta_z\leq h_0)\,,
\end{align}
where $v_{e;n,m,p}^\prime$ is the value of the fluctuation in the bin labeled with $(n,m,p)$. The covariance function $\mathcal C(\mathbf{x}- \mathbf{x^\prime})$ is defined as in Eq. \eqref{covfuncdef}.
Because of the stationarity of the field, the covariance function only depends on the relative positions in the medium. Since the correlation is non-zero only within the same bin, we can write
\begin{align}
\mathcal C(\mathbf{x}- \mathbf{x^\prime}) = \sum_{n,m,p} \langle & v^{\prime^2}_{e;n,m,p} \mathbb{I}_{\delta_x}(n) \mathbb{I}_{\delta_y}(m) \mathbb{I}_{\delta_z}(p) \nonumber \\
&\times \mathbb{I}_{\delta^\prime_x}(n) \mathbb{I}_{\delta^\prime_y}(m) \mathbb{I}_{\delta^\prime_z}(p) \rangle \,.
\end{align}
where the primed deltas refer to the point $\vx^\prime$ and the indicator functions are $1$ if the point is within the bin and $0$ otherwise.
The ensemble averaging is performed by integrating over the uniformly
distributed variables $\delta_i$ and $\delta^\prime_i$, with $i = x,
y, z$, as well as over the bins sizes $\ell$. By defining
$\boldsymbol\Delta\mathbf x = \vx - \vx^\prime$, the explicit
calculation leads to
\begin{align}
&\mathcal C(\mathbf{x}- \mathbf{\vx^\prime}) = \sigma_v^2 \int\limits_{\lvert \Delta x \rvert }^\infty d\ell \, p_\ell(\ell) \left(1 - \frac{ \lvert \Delta x \rvert }{\ell} \right) \nonumber \\
& \times \left(1 - \frac{ \lvert \Delta y \rvert }{d_0} \right)
  \left(1 - \frac{ \lvert \Delta z \rvert }{h_0} \right) 
, \label{covfunc}
\end{align}
and $0$ for $|\Delta y| > d_0$ or $|\Delta z| > h_0$. 
where $\sigma_v^2 = \mathcal C(\mathbf 0)$ is the variance of the Eulerian velocity. 
The correlation function is defined by 
$\mathscr{C}(\vx) = \mathcal C(\vx)/\sigma_v^2$.
By substituting the definition into Eq. \eqref{covfunc}, we observe that the covariance function can be factorized into
\begin{align}
\mathscr{C}(\vx) = \mathscr{X}(x)\mathscr{Y}(y)\mathscr{Z}(z)\, ,
\end{align}
where 
\begin{align}
\mathscr{X}(x) &= \int\limits_{\lvert x \rvert }^\infty d\ell \, p_\ell(\ell) \left(1 - \frac{ \lvert x \rvert }{\ell} \right) \nonumber \\ 
\mathscr{Y}(y) &= \left(1 - \frac{ \lvert y \rvert }{d_0} \right) H(d_0 -
  |\Delta y|)\nonumber \\
\mathscr{Z}(z) &= \left(1 - \frac{ \lvert z \rvert }{h_0} \right)
                 H(h_0 - |\Delta z|)\nonumber
\end{align}
represent the correlation functions in the $x$, $y$ and $z$ directions, respectively.

%%%%%%%%%%%%%%%%%%%%%%%%%%%%%%%%%%%%%%%%%%%%%%%%%%%%%%
\section{Spatial moments} \label{app:momentsinter}
%%%%%%%%%%%%%%%%%%%%%%%%%%%%%%%%%%%%%%%%%%%%%%%%%%%%%%
In the following, we derive expressions~\eqref{m1hat}-\eqref{m2hat} for the
first and second displacement moments and the asymptotic scalings of
the mean and variance. To this end, we define the $j$th moments of
$\Phi(x,t)$ and $P(x,t)$ as 
\begin{align}
\mu_j^{(\Phi)}(t) &= \int dx^j \Phi(x,t) 
\\
\mu_j^{(P)}(t) &= \int dx^j P(x,t) .
\end{align}
%
%%%%%%%%%%%%%%%%%%%%%%%%%%%%%%%%%%%%%%%%%%%%%%%%%%%%%%
\subsection{Derivation of mean and variance}
%%%%%%%%%%%%%%%%%%%%%%%%%%%%%%%%%%%%%%%%%%%%%%%%%%%%%%
The relationship between the particle density with and without interpolation is given in the Fourier and Laplace space by Eq. \eqref{cstar}. By substituting this expression into Eq. \eqref{moments1},
for the Laplace transform of the first and second moment of the spatial density $c(x,t)$ we get
\begin{align}
m_1^\ast (\lambda) &=\frac{\lambda}{1-\psi^\ast(\lambda)} \left[\mu^{(\Phi)}_1(\lambda)\mu_0^{(P)}(\lambda)\right. \nonumber \\
& \left.+\mu^{(P)}_1(\lambda)\mu_0^{(\Phi)}(\lambda)  \right] \label{m1appendix} \\
m_{2}^\ast (\lambda) &= \frac\lambda{1-\psi^\ast(\lambda)}\left[ \mu^{(\Phi)}_{2} (\lambda)\mu_0^{(P)}(\lambda)  \right.\nonumber \\
& + \left.2\mu^{(\Phi)}_{1} (\lambda)\mu^{(P)}_{1} (\lambda) + \mu^{(P)}_2(\lambda)\mu_0^{(\Phi)}(\lambda)\right] \,, \label{m2appendix}
\end{align}
where the $\mu$s are the moments of $P$ and $\Phi$ in Laplace space. The sub-index denotes the order of the moment and the super-index indicates the distribution. 
Notice that the zero-th order moment of $P(x,t)$ in Laplace space is given by 
$ \mu_0^{(P)}(\lambda) = \tilde{P}^\ast(k=0, \lambda)$. By using Eq. \eqref{Pstar} and by assuming an instantaneous injection of particles at $t=0$, we get
\begin{align} 
 \mu_0^{(P)}(\lambda) = \lambda^{-1}\,. \label{mu0P}
\end{align}
Analogously, the first and the second moments of $P(x,t)$ are calculated by applying the expressions for the moments \eqref{moments1} to the distribution of  Eq. \eqref{Pstar}.
Thus, we get 
\begin{align} 
&\mu_1^{(P)}(\lambda) = \frac{\mu_1^\ast(\lambda)}{\lambda[1-\psi^\ast(\lambda)]} \label{mu1P}\\
&\mu_{2}^{(P)}(\lambda) = \frac{1}{1-\psi^\ast(\lambda)}\left[2\mu_1^\ast(\lambda) \mu_1^{(P)}(\lambda) + \frac{\mu_{2}^\ast(\lambda)}{\lambda}\right]\,. \label{mu2P}
\end{align}
The zero-th moment of $\Phi(x,t)$ is defined as the integral of the distribution over the spatial domain. Integrating Eq. \eqref{Phifinal} yields $\mu^{(\Phi)}(t) = \int\limits_t^\infty \psi(\tau)d\tau$, whose Laplace
transform reads
\begin{align}\label{mu0Phi}
 \mu_0^{(\Phi)}(\lambda) = \frac{1- \psi^\ast(\lambda)}{\lambda}\,.
\end{align}
Finally, the first moment of $\Phi$ is
\begin{align}
 \mu^{(\Phi)}_{1} (\lambda)  =\frac{1}{\lambda^2}\int\limits_0^\lambda d\lambda^\prime \, \mu^\ast_{1} (\lambda^\prime) - \frac{1}{\lambda}\mu_1^\ast (\lambda) \,,  \label{mu1Phi}
\end{align}
while the second moment is given by
\begin{equation}\label{mu2Phi}
\mu^{(\Phi)}_{2}(\lambda) = \frac{2}{\lambda^3}\int\limits_0^\lambda d\lambda^\prime \lambda^\prime \mu^\ast_{2}(\lambda^\prime) - \frac{\mu^\ast_{2}(\lambda)}{\lambda} \,.
\end{equation}
By substituting Eqs. \eqref{mu0P}, \eqref{mu1P}, \eqref{mu0Phi} and \eqref{mu1Phi} into Eq. \eqref{m1appendix} we get Eq. \eqref{m1hat} for the Laplace transform of the first moment of particle density. %following expression for the Laplace transform of the first moment of
Analogously, by substituting Eqs. \eqref{mu0P}, \eqref{mu1P}, \eqref{mu2P}, \eqref{mu0Phi}, \eqref{mu1Phi} and \eqref{mu2Phi} into Eq. \eqref{m2appendix}, we get Eq. \eqref{m2hat}
for the Laplace transform of the second moment.
%
%
%%%%%%%%%%%%%%%%%%%%%%%%%%%%%%%%%%%%%%%%%%%%%%%%%%%%%
\subsection{Asymptotic scalings}\label{app:momentscalc}
%%%%%%%%%%%%%%%%%%%%%%%%%%%%%%%%%%%%%%%%%%%%%%%%%%%%%%
% 
In this section we derive explicitly the asymptotic scalings of the first and second centered moment of particle density for each scenario presented in Sect. \ref{section:transpbehaviour}.
%%%%%%%%%%%%%%%%%%%%%%%%%%%%%%%%%%%%%%%%%%%%%%%%%%%%%%%%%%%%%%%%%%%%%%%%%%%%%%%%
\subsubsection{Distribution-induced anomalous diffusion}\label{app:momentscalc1}
%%%%%%%%%%%%%%%%%%%%%%%%%%%%%%%%%%%%%%%%%%%%%%%%%%%%%%%%%%%%%%%%%%%%%%%%%%%%%%%%
We first show that $\psi(t)$ scales asymptotically as $t^{-1-\beta}$. By definition, $\psi(t) = \int_{-\infty}^\infty dx \psi(x,t)$. By using Eq. \eqref{psifinal}, the distributions of bins lengths
\eqref{p_ell} and the distribution of velocities \eqref{power:v}, we get that the $j$-th moment of $\psi(x,t)$ scales at long times as
%
% %
\begin{align}\label{psioftapp2}
\mu_j(t) \propto t^{-1-\beta}\int\limits_0^\infty dx \, x^{j+\beta} \euler^{-x/\ell_0},
\end{align}
where $\mu_0(t) = \psi(t)$.
For $t\to\infty$ the integral converges to $\Gamma(\beta+j+1)$. Thus, from Eq. \eqref{psioftapp2} we conclude that 
asymptotically $\mu_j(t)\propto t^{-1-\beta}$ for $j=0,1,2$. 
By making use of Tauberian theorems, we obtain that the Laplace transform of quantities that scale asymptotically as $t^{-1-\beta}$ behaves for small $\lambda$ as $1-a_1 \lambda^\beta$ for $\beta \in (0,1)$
and as $1-a_1\lambda +a_2\lambda^\beta$ for $\beta \in (1,2)$. Finally, if $\beta = 1$, the scaling is $1-a_1\lambda +a_3\lambda\ln\lambda$.
Thus, we get
\begin{align} \label{scalings1}
\mu_j^\ast(\lambda) \propto
\begin{cases}
 1 - a_1\lambda^\beta   & \beta\in(0,1) \\
 1-a_1\lambda +a_3\lambda\ln\lambda & \beta = 1 \\
 1 - a_1 \lambda + a_2\lambda^\beta  & \beta\in(1,2).
\end{cases}
\end{align}
The real coefficients $\{a_i\}_{i=1,..,3}$ depend on the specific distribution.  
% %
By substituting the corresponding scalings into Eq. \eqref{m1hat} and by taking the leading orders in $\lambda$ we get the asymptotic scalings in the Laplace domain of the first moment of particle
density 
\begin{align}
m_1^\ast (\lambda) \propto 
\begin{cases}
 \lambda^{-1-\beta} & \beta \in (0,1) \\
 \frac{\lambda^{-2}}{\ln \lambda} & \beta = 1\\
 \lambda^{-2} & \beta \in (1,2].
\end{cases}
\end{align}
The asymptotics in real time are obtained by the application of the Tauberian theorems, which provides the results listed in Table \ref{tableDIAD}. %
Analogously, we calculate the second moment by substituting the scalings of Eq. \eqref{scalings1} into Eq. \eqref{m2hat} and we get, in Laplace space
\begin{align}
m_2^\ast (\lambda) \propto 
\begin{cases}
 \lambda^{-1-2\beta} & \beta \in (0,1) \\
 \frac{\lambda^{-3}}{\ln^3 \lambda} & \beta = 1\\
 \lambda^{-3}+\lambda^{\beta-4} & \beta \in (1,2].
\end{cases}
\end{align}
The application of the Tauberian theorems provides the following scalings in the time domain
\begin{align}
m_2 (t) \propto 
\begin{cases}
 t^{2\beta} & \beta \in (0,1) \\
 \frac{t^2}{\ln^3 t} & \beta = 1\\
 t^2 + t^{3-\beta} & \beta \in (1,2].
\end{cases}
\end{align}
Recall that the second centered moment is given by $\kappa(t) = m_2(t) - m_1^2(t)$. By taking the leading orders in $t$, we obtain the results that are listed in Table \ref{tableDIAD}.
\begin{table}[h]
\begin{center}
\begin{tabular} {c|c|c|c} 
 & $\beta\in(0,1)$ &$\beta =1$ & $\beta\in(1,2]$\\ \hline \\
$m_1(t)$ & $t^\beta$ & $\frac{t}{\ln(t)} $& $t$ \\
$\kappa(t)$ & $t^{2\beta}$ & $\frac{t^2}{\ln^3(t)}$ & $t^{3-\beta}$ \\
% $\kappa_{22}(t)$ & $t^{\beta}$ & $\frac{t}{\ln(t)}$ & $t$ \\
\end{tabular}
\caption{Distribution-induced anomalous diffusion: asymptotic scalings of first moment and variance.}\label{tableDIAD}
\end{center}
\end{table}
%
%%%%%%%%%%%%%%%%%%%%%%%%%%%%%%%%%%%%%%%%%%%%%%%%%%%%%%%%%%%%%%%%%%%%%%%%%%%%
\subsubsection{Correlation-induced anomalous diffusion}\label{app:momentscalc2}
%%%%%%%%%%%%%%%%%%%%%%%%%%%%%%%%%%%%%%%%%%%%%%%%%%%%%%%%%%%%%%%%%%%%%%%%%%%%
As we did in the previous section, we start by deriving the scaling of $\psi(t)$. By using the PDF of transition times and lengths of Eq. \eqref{psifinal}, the distribution of bins sizes \eqref{power_ell}
and the velocity PDF \eqref{exp:v}, we get that the $j$-th moment of $\psi(x,t)$ is given by
\begin{align}\label{psioftappcorr1}
% \psi(t) = \frac{\alpha \ell_0^\alpha q t^{-\frac{1}{2}}}{\sqrt{2D}} \int \limits_{\ell_0}^\infty dx_1 x_1^{-2-\alpha} \euler^{-\frac{qt}{x_1}}\int\limits_{-\infty}^\infty dx_2 p_\eta\left(\frac{x_2}{\sqrt{2Dt}} 
% \right) \,.
\mu_j(t) = &\frac{\alpha\ell_0^\alpha }{t\sqrt{2\pi\sigma_e^2}}\int\limits_{\ell_0}^\infty dx \, x^{j-1-\alpha} \nonumber \\
&\times \exp{\left( -\frac{[\ln{(x/t)}-\mu_s]^2}{2\sigma_e^2}  \right)}.
\end{align}
We introduce the change of variable $y = x/t$. With this substitution, Eq.
\eqref{psioftappcorr1} simplifies to
\begin{align}\label{psioftappcorr2}
\mu_j(t) = &\frac{\alpha\ell_0^\alpha t^{j-1-\alpha}}{\sqrt{2\pi\sigma_e^2}}\int \limits_{\ell_0/t}^\infty dy \, y^{j-1-\alpha} \nonumber \\
& \times \exp{\left[-\frac{(\ln y - \mu_s)^2}{2\sigma_e^2} \right]}. 
\end{align}
For $t\to\infty$ the integral converges to a constant value. Thus,
by applying Tauberian theorems in the long time limit we get for $j=0$ 
\begin{align}\label{scalings2psi}
 \psi^\ast(\lambda) \propto 
 \begin{cases}
1-a_1\lambda^{\alpha} & \alpha \in (0,1) \\
 1 - a_1\lambda + a_3\lambda \ln \lambda & \alpha = 1 \\
1-a_1\lambda+ a_2\lambda^\alpha & \alpha\in(1,2) \,.
\end{cases}
\end{align}
%
% %
Analogously, we find that the first moment of $\psi(x,t)$ scales in Laplace space as
\begin{align} \label{scalings2mu1}
 \mu_1^\ast (\lambda) \propto
 \begin{cases}
    \lambda^{\alpha-1} & \alpha \in (0,1) \\
    1 + a_3\ln \lambda        & \alpha = 1 \\
    1-a_1\lambda^\alpha & \alpha\in(1,2) ,
 \end{cases}
\end{align}
while for the second moment we get
\begin{align} \label{scalings2mu2}
 \mu_2^\ast (\lambda) \propto
 \begin{cases}
    \lambda^{\alpha-2} & \alpha \in (0,1] \\
    \lambda^{\alpha-1} & \alpha\in(1,2).
 \end{cases}
\end{align}
By substituting the scalings of Eqs. \eqref{scalings2psi} and \eqref{scalings2mu1} into Eq. \eqref{m1hat}, we find
 \begin{align}
m_1^\ast (\lambda) \propto 
\begin{cases}
 \lambda^{-2} & \alpha \in (0,1) \\
 \frac{\lambda^{-2}}{\ln \lambda} & \alpha = 1\\
 \lambda^{-2} & \alpha \in (1,2].
\end{cases}
\end{align}
By applying the Tauberian theorems, we find the scalings in the time domain that are listed in Table \ref{table:exptheta}.
The scalings of the second moments are obtained in Laplace space by substituting Eqs. \eqref{scalings2psi},\eqref{scalings2mu1} and \eqref{scalings2mu2} into Eq. \eqref{m2hat}, which gives
 \begin{align}
m_2^\ast (\lambda) \propto 
\begin{cases}
 \lambda^{-3} & \alpha \in (0,1) \\
 \frac{\lambda^{-3}}{\ln \lambda} & \alpha = 1\\
 \lambda^{-3} + \lambda^{\alpha-4} & \alpha \in (1,2].
\end{cases}
\end{align}
The application of Tauberian theorems provides the following scalings in the time domain
\begin{align}
m_2 (t) \propto 
\begin{cases}
 t^{2} & \alpha \in (0,1) \\
 \frac{t^2}{\ln t} & \alpha = 1\\
 t^2 + t^{3-\alpha} & \alpha \in (1,2].
\end{cases}
\end{align}
Finally, by using these scalings for the calculation of the second centered moment and by taking the leading orders in $t$, we get the scalings that are listed in table \ref{table:exptheta}.
\begin{table}[h]
\begin{center}
\begin{tabular} {c|c|c|c} 
 & $\alpha\in(0,1)$ & $\alpha = 1$ & $\alpha\in(1,2]$\\ \hline \\
% $m_1(t)$ & $t$ & $\frac{t}{\ln t} $& $t$ \\
$m_1(t)$ & $t$ & $\frac{t}{\ln t} $& $t$ \\
$\kappa(t)$ & $t^{2}$ &$\frac{t^2}{\ln t} $& $t^{3-\alpha}$ \\
\end{tabular}
\caption{Correlation-induced anomalous diffusion: asymptotic scalings of first moment and variance.}\label{table:exptheta}
\end{center}
\end{table}

%%%%%%%%%%%%%%%%%%%%%%%%%%%%%%%%%%%%%%%%%%%%%%%%%%%%%%%%%%%%%%%%%%%%%%%%%%%%%%%%%%%%%%%%%%%%%%
\subsubsection{Anomalous diffusion induced by distribution and correlation} \label{momentscalc3}
%%%%%%%%%%%%%%%%%%%%%%%%%%%%%%%%%%%%%%%%%%%%%%%%%%%%%%%%%%%%%%%%%%%%%%%%%%%%%%%%%%%%%%%%%%%%%%
In this scenario the distribution of step lengths is given by \eqref{power_ell}, while the distribution of velocities is \eqref{power:v}. By substituting these expressions into Eq. \eqref{psifinal} and by 
calculating the spatial moments, we get 
\begin{equation}\label{psioftappDC2}
\mu_j(t) \propto  t^{-1-\beta} \int \limits_{\ell_0}^\infty dx \, x^{j+\beta-\alpha-1}\exp{\left(-\frac{x}{v_\textrm{max}t}\right)} \,.
\end{equation}
By using the change of variable $y = \frac{x}{v_\textrm{max}t}$ we get
\begin{align}
\mu_j(t) \propto t^{j-\alpha-1}\int\limits_{\frac{\ell_0}{v_{\textrm{max}}}}^\infty dy \, y^{j+\beta-\alpha-1}\exp{(-y)}.
\end{align}
For different values of $\alpha$ and $\beta$ very different asymptotic behaviors arise. In particular, we get
\begin{equation}\label{scalingpsit}
\mu_j(t)\propto
\begin{cases}
t^{j-1-\alpha} & \alpha < \beta+j \\
t^{-1-\omega}\ln t & \alpha = \beta +j \\
t^{-1-\beta} & \alpha > \beta +j,
\end{cases}
\end{equation}
where $\omega = \min(\alpha,\beta)$.
Recall that the range of the parameters does not allow the possibility $\alpha \ge \beta+2$. Therefore, a unique expression for the scaling of the second moment of $\psi(x,t)$ is found. Namely, we get $\mu_2(t) \propto t^{1-\alpha}$. 
% %
In the following, we will treat four different scenarios, neglecting the cases in which $\alpha = \beta$ because of their unlikelihood. Nevertheless, those cases are reported in Table \ref{table:parepare1} for completeness.
%%%%%%%%%%%%%%%%%%%%%%%%
\paragraph{Case $\alpha,\beta \in (0,1); \alpha\neq\beta$} By using Tauberian theorems into Eq. \eqref{scalingpsit}, we find that the distribution of transition times admits the following expansion for large times in Laplace domain
\begin{equation}\label{psialpha}
\psi^\ast(\lambda)\propto 1-a_1\lambda^{\omega}.  
\end{equation}
The first moment, given by Eq. \eqref{scalingpsit} for $j=1$, scales in Laplace space as $\mu_1^\ast(\lambda) \propto \lambda^{\alpha-1}$, while the second moment scales as $\mu_2^\ast(\lambda) \propto \lambda^{\alpha-2}$.
By substituting the so-obtained scalings into Eq. \eqref{m1hat} and \eqref{m2hat}, we get for the first and the second moment of particles density
\begin{align}
m_1^\ast(\lambda) \propto \lambda^{-1-\nu} & &m_2^\ast(\lambda) \propto \lambda^{-1-\mu} + \lambda^{-1-\epsilon},
\end{align}
where $\nu = \min(1, \beta - \alpha +1)$ and $\epsilon = \min(2, 2 + \beta -\alpha)$. By applying the Tauberian theorems, we find the scalings summarized in Table \ref{table:parepare2}.
%
%%%%%%%%%%%%%%%%%%%%
\paragraph{Case $\alpha,\beta \in (1,2); \alpha\neq\beta$} In this case, for long times Eq. \eqref{scalingpsit} can be expanded in Laplace space as 
\begin{equation}
\psi^\ast(\lambda) \propto 1 - a_1\lambda + a_2 \lambda^\omega,
\end{equation}
while the first and the second moments of $\psi(x,t)$ scale as  $\mu_1^\ast(\lambda) \propto \lambda^{\alpha-1}$ and as $\mu_2^\ast(\lambda) \propto \lambda^{\alpha-2}$, respectively. By applying the usual methodology, we get
\begin{align}
m_1^\ast(\lambda) \propto \lambda^{-2} & & m_2^\ast(\lambda) \propto \lambda^{-3} + \lambda^{\omega-4},
\end{align}
After applying the Tauberian theorems and by using $\kappa(t) = m_2(t) - m_1^2(t)$, for the first and second centered moment of particle displacements, we get the scalings listed in table \ref{table:parepare2}.
%%%%%%%%%%%%%%%%%%%%%%%%%%%
\paragraph{Case $\alpha\in(0,1), \beta \in (1,2)$}  For this scenario, by using Eq. \ref{scalingpsit} and Tauberian theorems, we get again the scaling of Eq. \eqref{psialpha} for $\psi^\ast(\lambda)$ while the first and the second moment scale as $\mu_1^\ast(\lambda) \propto \lambda^{\alpha-1}$ and $\mu_2^\ast(\lambda) \propto \lambda^{\alpha-2}$, respectively (note that here $\alpha < \beta +1$).
In this case, we get from Eqs. \eqref{m1hat} and \eqref{m2hat}
\begin{align}
m_1^\ast(\lambda) \propto \lambda^{-2} & & m_2^\ast(\lambda) \propto \lambda^{-3}. 
\end{align}
The corresponding scalings in time domain are listed in Table \ref{table:parepare3}.
%%%%%%%%%%%%%%%%%%%%%%%%%%%
\paragraph{Case $\alpha\in(1,2), \beta \in (0,1)$} In this last scenario, we get from Eq. \eqref{scalingpsit} and from the Tauberian theorems
\begin{equation}
\psi^\ast(\lambda) \propto 1 -a_1\lambda^\beta,
\end{equation}
while the second moment of $\psi(x,t)$ scales as $\mu_2^\ast(\lambda) \propto \lambda^{\alpha-2}$, respectively. The first moment, given by Eq. \eqref{scalingpsit} for $j=1$, scales as
\begin{align}
\mu_1^\ast(\lambda) \propto 
\begin{cases}
 \lambda^{\alpha-1} & \alpha < \beta +1 \\
 1-a_1\lambda^\beta & \alpha > \beta +1. 
 \end{cases}
\end{align}
By substituting the scalings here derived into Eqs. \eqref{m1hat} and \eqref{m2hat} we get in Laplace space
\begin{align}
m_1^\ast(\lambda) \propto \lambda^{-1-\beta} & & m_2^\ast(\lambda) \propto \lambda^{-1-2\beta}. 
\end{align}
The scalings in time domain are obtained through the application of Tauberian theorems and are listed in Table \ref{table:parepare3}.

\begin{table}[h]
\begin{center}
\begin{tabular} {c|c|c|c} 
 & $\alpha=\beta\in(0,1)$ & $\alpha = \beta = 1 $&  $\alpha=\beta\in(1,2)$  \\ \hline \\
$m_1(t)$ & $\frac{t}{\ln(t)}$ & $\frac{t}{\ln^2 t} $ & $t$ \\
$\kappa(t)$ & $\frac{t^2}{\ln(t)}$& $\frac{t^2}{\ln^2 t} $ & $t^{3-\alpha}\ln(t)$\\
\end{tabular}
\caption{Anomalous transport induced by distribution and correlation: asymptotic scaling of the first moment and variance.}\label{table:parepare1}
\end{center}
\end{table}
\begin{table}[h]
\begin{center}
\begin{tabular} {c|c|c} 
 &  $\alpha\neq\beta$, $\alpha,\beta\in(0,1)$ & $\alpha\neq\beta$, $\alpha,\beta\in(1,2)$ \\ \hline \\
$m_1(t)$ & $t^\nu$ & $t$ \\
$\kappa(t)$ & $t^{\varepsilon}$ & $t^{3-\omega}$ \\
\end{tabular}
\caption{Anomalous transport induced by distribution and correlation: asymptotic scaling of the first moment and variance.}\label{table:parepare2}
\end{center}
\end{table}
\begin{table}[h]
\begin{center}
\begin{tabular} {c|c|c} 
 &  $\alpha\in (0,1), \beta\in (1,2)$ & $\alpha\in (1,2), \beta\in (0,1)$ \\ \hline \\
$m_1(t)$ & $t$ & $t^\beta$ \\
$\kappa(t)$ & $t^{2}$ & $t^{2\beta}$ \\
% $\kappa_{22}(t)$ & $t$ & $t^\beta$ \\
\end{tabular}
\caption{Anomalous transport induced by distribution and correlation: asymptotic scaling of the first moment and variance.}\label{table:parepare3}
\end{center}
\end{table} 
%
%\bibliographystyle{ieeetr}
%\bibliography{CoupledCTRW}

\end{document}